\documentclass[a4paper]{aa}
\usepackage{txfonts,graphicx,natbib}
\bibpunct{(}{)}{;}{a}{}{,}

\def\Teff{$T_{\mathrm{eff}}$}
\def\logg{\ensuremath{\log g}}
\def\vmic{$\upsilon_{\mathrm{mic}}$}
\def\vmac{$\upsilon_{\mathrm{macro}}$}
\def\vsini{\ensuremath{{\upsilon}\sin i}}
\def\kms{$\mathrm{km\,s}^{-1}$}
\def\ms{$\mathrm{m\,s}^{-1}$}
\def\exc{$\chi_{\mathrm{excit}}$}

\def\nlte{non-LTE}
\def\llm{{\sc LLmodels}}
\def\hfs{{\it hfs}}

\def\width{{\sc WIDTH9}}
\def\synth{{\sc SYNTH3}}
\def\vald{{\sc VALD}}

\begin{document}
\title{An In-Depth Spectroscopic Analysis of the Blazhko Star RR Lyr
\thanks{Data obtained with the 2.7-m telescope at McDonald Observatory, Texas, US}
} 
\subtitle{I. Characterisation of the star: abundance analysis and fundamental parameters } 
\author{K. Kolenberg\inst{1} \and
       L. Fossati\inst{2} \and 
	D. Shulyak\inst{3} \and
	H. Pikall\inst{1}    \and
	T.G. Barnes\inst{4}  \and
	O. Kochukhov\inst{5}  \and
	V. Tsymbal\inst{6}
	}
\institute{
	Institut f\"ur Astronomie, Universit\"{a}t Wien, 
	T\"{u}rkenschanzstrasse 17, 1180 Wien, Austria.\\
	\email{katrien.kolenberg@univie.ac.at, holger.pikall@univie.ac.at} 
	\and
	Department of Physics and Astronomy, Open University, Walton Hall, 
	Milton Keynes, MK7 6AA, UK.\\
	\email{l.fossati@open.ac.uk}
	\and
	Institute of Astrophysics, Georg-August-University, 
	Friedrich-Hund-Platz 1, D-37077, G\"ottingen, Germany.\\
	\email{denis.shulyak@gmail.com} 
	\and
	The University of Texas at Austin, McDonald Observatory, 1 
	University Station, C1402, Austin, Texas, 78712-0259, USA.\\
	\email{tgb@astro.as.utexas.edu}
	\and
	Department of Astronomy and Space Physics, Uppsala University, 
	751 20, Uppsala, Sweden.\\
	\email{oleg@astro.uu.se}
	\and  
	Tavrian National University, Vernadskiy's Avenue 4, Simferopol, 
	Crimea, 95007, Ukraine.\\
	\email{vadim.tsymbal@gmail.com}
	} 
\date{} 
\abstract
{}
{The knowledge of accurate stellar parameters is a keystone in several fields
of stellar astrophysics, such as asteroseismology and stellar evolution. 
Although the fundamental parameters can be derived both from spectroscopy and multicolour 
photometry, the results obtained are sometimes affected by systematic 
uncertainties. In this paper, we present a self-consistent spectral 
analysis of the pulsating star RR~Lyr, which is the primary target for our 
study of the Blazhko effect.}
{We used high-resolution and high signal-to-noise ratio spectra to carry 
out a consistent parameter determination and abundance analysis for RR~Lyr. 
The \llm\ code was employed for model atmosphere calculations, while \synth\ 
and \width\ codes were used for line profile calculation and LTE abundance 
analysis.  We provide a detailed description of the methodology 
adopted to derive the fundamental parameters and the abundances. 
Stellar pulsation reaches high amplitudes in RR~Lyrae stars, and as a consequence the stellar parameters vary significantly over the pulsation cycle.  The abundances of the star, however, are not expected to change. From a set of available high-resolution spectra of RR~Lyr we selected the phase of maximum radius, at which the spectra are least disturbed by the pulsation.
Using the abundances determined at this phase as a starting point, we expect to obtain a higher accuracy in the fundamental parameters determined at other phases. }
{The set of fundamental parameters obtained in this work fits the observed spectrum accurately. 
Through the abundance analysis, we find clear indications for a depth-dependent microturbulent velocity, 
that we quantified. 
}
{We confirm the importance of a consistent analysis of relevant spectroscopic 
features, application of advanced model atmospheres, and the use of up-to-date 
atomic line data for the determination of stellar parameters. These 
results are crucial for further studies, e.g., detailed theoretical modelling 
of the observed pulsations.
}
\titlerunning{}
\authorrunning{K.~Kolenberg et al.}
\maketitle
\section{Introduction}\label{introduction}
The modelling of pulsational signals requires the knowledge of stellar
parameters and primarily accurate values of the effective temperature (\Teff) 
and metallicity ($Z$). The determination of fundamental parameters can 
be performed by different methods
(some examples for stars from B- to G- type are Fuhrmann et al. 1997, Przybilla et al. 2006, and Fossati et al. 2009)
that do not always lead to consistent results. Thus, it is important to 
choose a methodology that allows us to constrain the parameters of the star from 
the available observables (usually photometry and spectroscopy) in the most 
robust and reliable way.

RR~Lyr is the prototype and eponym of its class of pulsating stars. RR Lyrae stars play a crucial role as distance indicators.  Their evolutionary stage 
(He burning in core, H burning in shell) makes them useful tracers of galactic history.  These classical pulsators display radial oscillations (the simplest type of pulsation) with large amplitudes which makes them useful touchstones for theoretical modelling. 
RR~Lyr is one of the best studied stars of its class.  Almost a century ago Shapley (1916) discovered that it shows a strong Blazhko effect, i.e. a 
(quasi-)periodic modulation of its light curve shape in amplitude and phase. 
The Blazhko effect in RR~Lyr has been closely followed over the past century, and changes have been reported both in the strength and the duration of its Blazhko cycle (Szeidl 1988, Kolenberg et al. 2006). Some well-studied stars even show multiple (variable) modulation periods (see, e.g., LaCluyz\'e et al. 2004).
Despite numerous attempts to model the phenomenon, the Blazhko effect has eluded a satisfactory explanation so far.  
Recently obtained high-precision photometry from ground-based or space-borne precise instruments also indicate that Blazhko modulation may be a much more common phenomenon than initially thought: as many as half of galactic RRab stars may be modulated  (Jurcsik et al. 2009;  Szabo et al. 2009; Kolenberg et al. 2010).  

In order to constrain the viable models for the Blazhko effect, it is vital to obtain accurate values for the fundamental parameters (and their variations) for modulated and non-modulated RR Lyrae stars.  This has been the main motivation for the study presented in this article. 

RR~Lyr is the only star of its class to have a directly determined parallax, recently measured with the HST/FGS, by Benedict et al. (2002), to be $\pi$(FGS) = 3.82 $\pm$ 0.2 mas (d = 262 $\pm$ 13 pc). When a small ISM correction of $A_v$ = 0.07 is applied, this new distance results in an $M_{v}^{\rm RR} \simeq +0.61_{-0.11}^{+0.10}$ mag which corresponds to $\simeq 49 \pm 5 L_{\odot}$. 

Fundamental parameters of RR Lyr have been obtained by several authors with a variety of methods (e.g., Lambert et al. 1996; Manduca et al. 1981; Siegel 1982; for a summary see Kolenberg 2002). The published fundamental 
parameters of RR~Lyr display a considerable range both in \Teff\ and \logg\  due to the large pulsation amplitudes.
According to these analyses, RR~Lyr's  \Teff\ varies over its 13h36min pulsation cycle between approximately 6250 and 8000 K and its \logg\ between 2.5 and 3.8 (extreme values). 
Superposed on the large variation, the Blazhko cycle leads to an additional variation 
of the fundamental  parameters. Jurcsik et al. (2008) recently showed that also the {\it mean} properties of  modulated RR~Lyrae stars change over the Blazhko cycle. 
Element abundances of RR~Lyr were obtained previously by, e.g., Clementini et al. (1996), Lambert et al. (1996), and Takeda et al. (2006). 

The main goal of the present work is to perform a self-consistent atmospheric 
and abundance analysis of RR~Lyr that reproduces all of its photometric and 
spectroscopic data. Furthermore, we want to investigate the degree to which the derived 
fundamental parameters depend on the applied methods.
Considering the structure of the available models, especially the position of the 
convective zones and the zones of nuclear fusion, the measured abundances of 
the star are not expected to change over the pulsation (and the Blazhko) cycle.  Hence, if the abundances are accurately determined at one phase in the pulsation cycle, they can be of help to determine (or at least constrain) the fundamental parameters at other phases.   In this paper we also selected the optimal phase for determining the abundances of the star. This is the first of a series of planned papers devoted to a detailed spectroscopic study of RR~Lyr. In forthcoming papers we will discuss the spectral variations over the pulsation and Blazhko cycle of the star.

\section{Observations and spectral data reduction}\label{observations-sect}

A total of 64 spectra of RR Lyr were obtained between
June 26th and August 27th, 2004 with the Robert G. Tull Coud\'e
Spectrograph on the 2.7-m telescope of McDonald Observatory.  This
is a cross-dispersed \'echelle spectrograph yielding a two-pixel
resolving power $R \simeq 60 000$ for the configuration used here.
Table~\ref{observations} lists the observing time, exposure time, the S/N ratio
per resolution element for each acquired spectrum, and the phases
in the pulsation and Blazhko cycles. For the determination of the
phases we used the ephemerides derived by Kolenberg et al. (2006).
To minimize smearing of the spectral features by pulsation, each
spectrum was limited to an exposure time of 960 seconds. Two
spectra have shorter exposures as a result of being stopped due
to cloud. The signal-to-noise ratio (SNR) per resolution element
of the obtained spectra varies according to the brightness of the
star (given the fixed integration time) and the weather conditions
during the observation. Spectra ID319-328 were inadvertently
taken at the wrong blaze angle and thus have poor S/N ratios.

Bias frames and flat-field frames were obtained at the
start of each night, and Th-Ar spectra were observed frequently
during each night for calibration purposes.  The spectra were
reduced using the Image Reduction and Analysis Facility  
(IRAF{\footnote{IRAF {\tt (http://iraf.noao.edu/)} 
is distributed by the National Optical
Astronomy Observatory, which is operated by the Association of
Universities for Research in Astronomy (AURA) under cooperative
agreement with the National Science Foundation.},
Tody 1993).  Each spectrum, normalised by fitting a low order
polynome to carefully selected continuum points, covers the
wavelength range 3633-10849\,\AA, with several gaps between the
orders at wavelengths greater than 5880\,\AA.

\begin{table}[ht!]
\caption[ ]{Basic data of the observations of RR~Lyr.}
\label{observations}
\begin{center}
\scriptsize{
\begin{tabular}{l|c|c|c|c|c}
\hline
\hline
Spectrum ID & HJD $-$ & Pulsational & Blazhko & SNR per  & Exposure \\       
number   & 2453000 & phase       & phase   & pixel & time (s) \\
\hline
087	&	 183.7953   & 0.173 &  0.280 &  293 &   960.000    \\
088	&	 183.8147   & 0.207 &  0.281 &  250 &   960.000    \\
089	&	 183.8271   & 0.229 &  0.281 &  172 &   960.000    \\
091	&	 183.8441   & 0.260 &  0.281 &  149 &   960.000    \\
119	&	 184.7293   & 0.820 &  0.304 &  215 &   960.000    \\
120	&	 184.7434   & 0.846 &  0.305 &  253 &   960.000    \\
121	&	 184.7559   & 0.868 &  0.305 &  271 &   960.000    \\
122	&	 184.7683   & 0.890 &  0.305 &  194 &   960.000    \\
124	&	 184.7865   & 0.922 &  0.306 &  162 &   960.000    \\
125	&	 184.7988   & 0.943 &  0.306 &  126 &   960.000    \\
126	&	 184.8121   & 0.967 &  0.306 &  63  &   960.000    \\
158	&	 185.7401   & 0.604 &  0.330 &  101 &   960.000    \\
159	&	 185.7523   & 0.626 &  0.331 &  126 &   960.000    \\
160	&	 185.7645   & 0.647 &  0.331 &  171 &   960.000    \\
161	&	 185.7768   & 0.669 &  0.331 &  233 &   960.000    \\
163	&	 185.7937   & 0.698 &  0.332 &  232 &   960.000    \\
164	&	 185.8059   & 0.720 &  0.332 &  259 &   960.000    \\
165	&	 185.8182   & 0.741 &  0.332 &  113 &   960.000    \\
166	&	 185.8304   & 0.763 &  0.333 &  210 &   960.000    \\
168	&	 185.8465   & 0.792 &  0.333 &  228 &   960.000    \\
169	&	 185.8587   & 0.814 &  0.333 &  290 &   960.000    \\
170	&	 185.8709   & 0.834 &  0.334 &  236 &   960.000    \\
171	&	 185.8831   & 0.856 &  0.334 &  216 &   960.000    \\
173	&	 185.8989   & 0.885 &  0.334 &  236 &   960.000    \\
174	&	 185.9112   & 0.905 &  0.335 &  215 &   960.000    \\
175	&	 185.9234   & 0.928 &  0.335 &  233 &   960.000    \\
176	&	 185.9356   & 0.948 &  0.335 &  290 &   960.000    \\
178	&	 185.9512   & 0.977 &  0.336 &  357 &   960.000    \\
204	&	 186.7293   & 0.349 &  0.356 &  190 &   960.000    \\
205	&	 186.7425   & 0.372 &  0.356 &  138 &   960.000    \\
206	&	 186.7548   & 0.394 &  0.356 &  203 &   960.000    \\
207	&	 186.7670   & 0.416 &  0.357 &  46  &	960.000    \\
209	&	 186.7876   & 0.452 &  0.357 &  131 &   960.000    \\
210	&	 186.8007   & 0.475 &  0.358 &  77  &   960.000    \\
251	&	 187.7208   & 0.098 &  0.381 &  359 &   960.000    \\
252	&	 187.7330   & 0.120 &  0.382 &  319 &   960.000    \\
253	&	 187.7442   & 0.141 &  0.382 &  120 &   769.397    \\
255	&	 187.7632   & 0.173 &  0.382 &  124 &   960.000    \\
256	&	 187.7803   & 0.203 &  0.383 &  160 &   960.000    \\
257	&	 187.7923   & 0.226 &  0.383 &  87  &   797.949    \\
258	&	 187.8060   & 0.249 &  0.383 &  272 &   960.000    \\
{\bf 260}	&	 187.8226   & 0.278 &  0.384 &  322 &   960.000    \\
261	&	 187.8370   & 0.303 &  0.384 &  214 &   960.000    \\
262	&	 187.8504   & 0.327 &  0.385 &  288 &   960.000    \\
263	&	 187.8634   & 0.349 &  0.385 &  216 &   960.000    \\
319	&	 243.6989   & 0.854 &  1.824 &  27  &   960.000    \\
322	&	 243.7688   & 0.976 &  1.826 &  21  &   960.000    \\
323	&	 243.7832   & 0.002 &  1.826 &  18  &   960.000    \\
324	&	 243.7955   & 0.024 &  1.827 &  18  &   960.000    \\
326	&	 243.8123   & 0.053 &  1.827 &  21  &   960.000    \\
327	&	 243.8262   & 0.078 &  1.827 &  25  &   960.000    \\
328	&	 243.8395   & 0.101 &  1.828 &  13  &   960.000    \\
363	&	 244.6062   & 0.454 &  1.847 &  231 &   960.000    \\
364	&	 244.6201   & 0.478 &  1.848 &  251 &   960.000    \\
365	&	 244.6323   & 0.500 &  1.848 &  258 &   960.000    \\
366	&	 244.6445   & 0.521 &  1.848 &  238 &   960.000    \\
368	&	 244.6622   & 0.553 &  1.849 &  249 &   960.000    \\
369	&	 244.6744   & 0.574 &  1.849 &  181 &   960.000    \\
370	&	 244.6870   & 0.596 &  1.849 &  173 &   960.000    \\
371	&	 244.6992   & 0.618 &  1.850 &  171 &   960.000    \\
373	&	 244.7168   & 0.648 &  1.850 &  152 &   960.000    \\
374	&	 244.7290   & 0.670 &  1.851 &  122 &   960.000    \\
375	&	 244.7412   & 0.693 &  1.851 &  90  &   960.000    \\
376	&	 244.7535   & 0.713 &  1.851 &  69  &   960.000    \\
\hline											  
\end{tabular}}
\end{center}
\footnotesize{The first column 
shows the spectrum ID number and the second the Heliocentric Julian Date
(HJD-2453000) at the beginning of the exposure. The third and fourth columns 
show the pulsational and Blazhko phase respectively for each observation. The 
fifth column shows the SNR per pixel calculated at $\sim$5000\,\AA., while 
the sixth column shows the exposure time, in seconds, for each observation. 
The spectra number 253 and 257 have a lower exposure time, because the 
observation was stopped due to the presence of thick clouds. 
Spectrum number {\bf 260} is the one analyzed in detail in this paper.}

\end{table}



The normalisation of the hydrogen lines was a crucial point since we used them
as source for the derivation of the \Teff.
H$\alpha$ was not covered by our spectra, and the orders adjacent to H$\beta$ were affected by a spectrograph defect which hindered a proper normalization. We were able to perform a reliable normalisation of the H$\gamma$ line using the artificial flat-fielding technique described by Barklem et al. (2002).
This approach assumes that the relation between the blaze shapes of the different 
\'echelle orders is a smoothly changing function of the order number. On 
this basis one can establish the apparent blaze shapes by fitting polynomials 
to continuum points in several orders above and below the hydrogen line. On 
the subsequent step a 2-D polynomial surface is fitted to these empirical 
blaze functions, and the continuum in the orders containing the H$\gamma$ line 
is determined by interpolation.

This normalisation procedure was performed on two spectra taken near the optimal phase.
We used a surface fit to 3--4 orders on both sides 
of the broad hydrogen line to determine the continuum in the H$\gamma$ 
spectral orders. The accuracy of this normalisation technique is attested 
by a good agreement between the normalised overlapping H$\gamma$ 
profiles and by the lack of discrepancy between observations of RR~Lyr 
obtained at similar pulsation phases.

Simultaneously with the spectroscopic campaign, we obtained photometric 
data in Johnson $V$ through a multi-site campaign. The photometric data were 
published by Kolenberg et al. (2006). They were used for accurate determination 
of the pulsation frequencies and phases in the pulsation and Blazhko cycle.
\section{The models}

\subsection{The pulsation model}

In order to determine the dynamical properties of the atmosphere at the Ómost
quiescentÓ phases theoretically, we used the so-called ÒVienna
Nonlinear Pulsation CodeÓ.
For the unperturbed starting model we utilized the values of 
$0.65 M_{\odot}$ and $50 L_{\odot}$ together with $T_{\rm eff}=6600 $K   and 
a typical Pop II chemical composition of $Y=0.239$ and $Z=0.001$ which lead to 
a limit cycle with the observed period of pulsation.  The kinetic energy of the atmosphere (defined as the part from the photosphere 
where $\tau = 2/3$ to the outer 
boundary of our model, see Fig.~\ref{model-phase}) shows 2 local minima, where the first 
-- roughly at phase 0.35 -- corresponds to the phase of maximum photospheric radius. 
The flow ceases and the whole envelope starts to contract again to reach its minimal radius  
shortly after minimum light. Although velocities in the atmosphere are
low, we are aware that the atmosphere is not static at any point
during the starÕs pulsation, which is contrary to what is assumed in most
model atmosphere codes.  But there are phases where the dynamical effects are smaller. These are the phases we are interested in for abundance analysis.
Note that the pulsation model does not take into account the Blazhko modulation in the star.

\subsection{The model atmosphere}

To compute model atmospheres of RR~Lyr, we employed the \llm\ stellar 
model atmosphere code (Shulyak et al. 2004). For all the calculations, Local 
Thermodynamical Equilibrium (LTE) and plane-parallel geometry were assumed. 
Both these assumptions need to be discussed in our particular case.}

The {\it  LTE assumption} may be questionable
due to low plasma density and shock waves in the RR~Lyr atmosphere. To reduce the uncertainties
of spectroscopic analysis we implement our model atmospheres at phases where the
dynamical effects in the star's atmosphere are expected to be small. We refer the reader
to the next section for more explanations.  Ignoring LTE may also lead to systematic errors in the abundance
analysis. However, a detailed \nlte\ analysis is beyond the scope of the present paper. 

Atmospheres of giants are extended due to large radii and thus sphericity
effects may become important for the stellar atmosphere modelling. For instance,
based on detailed model atmosphere analysis Heiter \& Eriksson (2006) recommended
using spherically symmetric models for abundance analysis of stars having
$\log g \leqslant 2$ and $4000$~K~$\leqslant T_{\rm eff} \leqslant 6500$~K.
Taking into account the estimated gravitational acceleration of RR~Lyr of 
$\log g=2.4$ (see next section) one can expect the sphericity effects 
to be small enough to not
significantly influence the abundance analysis. Indeed, the estimated
errors of the Fe abundance based on theoretical \ion{Fe}{i} lines presented in
Heiter \& Eriksson (2006) do not exceed $0.1$~dex for the model with
$T_{\rm eff}=6500, \log g=1$.  A $\log g=2.4$ value for RR~Lyr thus 
justifies the use of {\it plane-parallel model atmospheres} for abundance analysis.

We used the \vald\ database 
(Piskunov et al. 1995; Kupka et al. 1999, Ryabchikova et al. 1999)
as a source of atomic 
line parameters for opacity calculations with the \llm\ code.
Finally, convection was implemented according to the 
Canuto \& Mazzitelli (1991a,b)
model of convection (see Heiter et al. (2002) for more details).

\section{The optimal phase}\label{right-phase}

During the pulsation cycle the spectral lines of RR~Lyr change dramatically. In
particular, when a shock wave passes through the atmosphere,  it is possible to observe line broadening, 
line doubling, line disappearance, and sometimes even line emission (Preston et al. 1965, Chadid \& Gillet 1996, Chadid et al. 2008, Preston 2009). 
This is obviously the sign of a very 
non-quiescent atmosphere, that in principle cannot be modelled with a static model 
atmosphere, such as ATLAS and \llm. In practice a model atmosphere code that 
is able to realistically model the atmosphere of a variable star such as 
RR~Lyr, given both its chemical and pulsational peculiarity, does not yet 
exist. Therefore we are forced to use a static model atmosphere.  
In order to analyse the star in the most consistent way, we decided to 
study RR~Lyr when its atmosphere is as close as possible to that of a 
non-variable star.

In the past, RR~Lyrae stars were always analysed through spectra obtained 
close to the phase of minimum light because it was believed that this 
was the phase at which the star's atmosphere is ``at its quietest". 
The adoption of 
static model atmospheres for the analysis of RR~Lyrae stars was then 
justified through showing that the results obtained at different phases were 
all in agreement with each other. 
At the ``quiescent phase'' we generally suppose that a) there are no shock-waves or
any other fast plasma motions in the atmosphere 
that distort the line profiles and b) pressure stratification is as close as possible
to its hydrostatic analog. 

Picturing a homogeneously pulsating sphere, the most quiescent phase is associated with both  
phases of minimum and maximum radius.
In case of an oversimplified pulsation model, these also correspond 
to the minimum and maximum light in the star's luminosity variation.
At these extreme positions in the pulsation, the atmosphere comes to a halt and, due to zero gas velocity, 
the pressure stratification at these phases is closest
to the hydrostatic case (but not necessarily the same!). Furthermore, since the 
plasma velocities are negligible, the kinetic energy should be zero as well. 
For RR~Lyrae stars, and especially those of type RRab (fundamental mode pulsators) with strongly nonlinear light curves (resembling a saw-tooth function), the phase of maximum light is short-lived and known to be accompanied by shock-waves.
Thus, in previous works the quiescent phase was associated with the phase of minimum light.
However, the assumption that all
atmospheric layers move synchronously as being rigidly bound to each other is wrong.
Realistic models show that at minimum radius, 
radiation is blocked in deeper layers and ready to migrate into the outer layers and 
accelerate both the photosphere and the atmosphere. 

\begin{figure}[ht]
\begin{center}
\includegraphics[height=120mm]{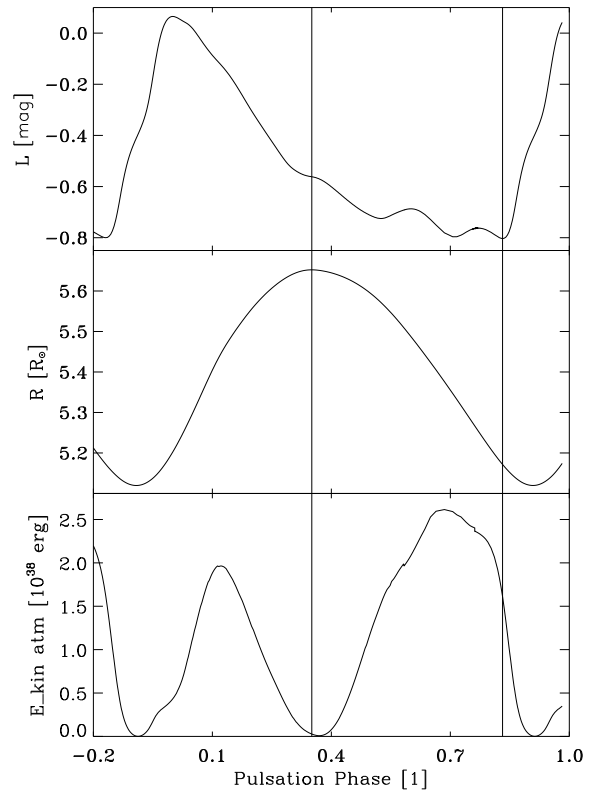}
\caption{RR~Lyr model bolometric luminosity (upper panel), 
photospheric radius (middle panel), and 
atmospheric kinetic energy (lower panel) as a function of the pulsational 
phase. As convention phase 0 corresponds to the maximum of the luminosity. The
two full vertical lines correspond to the phases of maximum radius (the phase we
declare as most quiescent) and of minimum light (usually adopted for the spectroscopic
analysis).}
\label{model-phase}
\end{center}
\end{figure}

Looking at both pulsational models and a large sample of stellar spectra,
obtained at different phases, we found that there is a
quiescent phase very close to the phase of maximum radius, where the 
radial velocity derived from the metallic lines corresponds to the stellar 
gamma velocity. On the light curve this corresponds to a phase on the descending branch of the light variation.
Figure~\ref{model-phase} shows the calculated bolometric light curve,the photospheric radius variation, and the atmospheric kinetic energy as a function of the pulsation phase. 
The two vertical lines correspond to the phases of maximum stellar radius and of minimum light.
As mentioned above the expansion of the model envelope is not homologous. 
We see different waves running in and out -- some even steepening to shock waves.

\begin{figure}[ht]
\begin{center}
\includegraphics[width=85mm, angle=0]{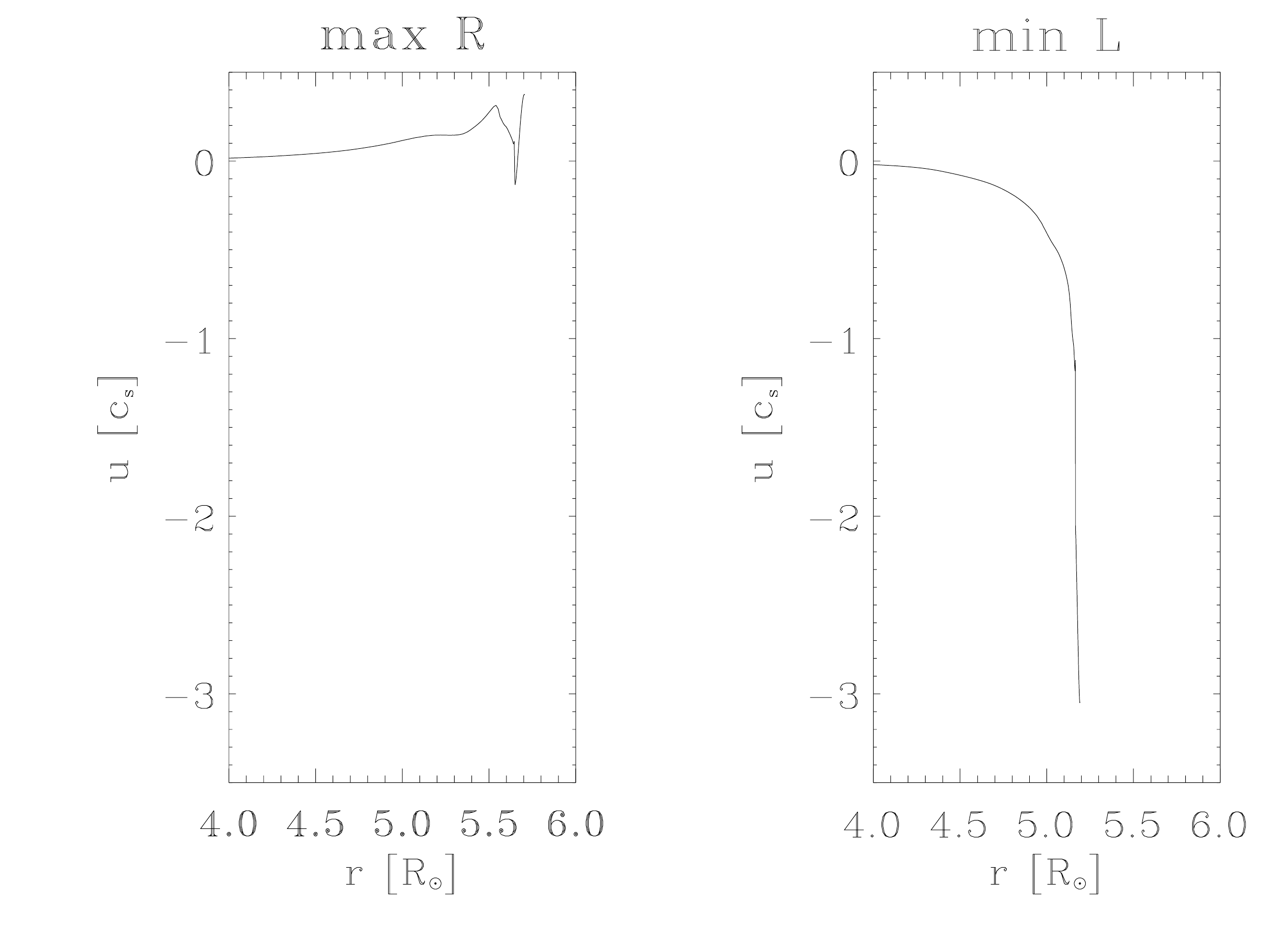}
\caption{Radial plots showing gas velocity $u$ in units of the sound speed (the so-called Mach number) shown for the phase of maximum radius (left) and minimum light (right).}
\label{comp}
\end{center}
\end{figure}

Minimum light is a less fortunate choice to obtain undistorted line profiles, as parts of the atmosphere still move with supersonic speed (e.g., Mach 3 in Fig.\,\ref{comp}). 
Fig.\,\ref{comp}  shows radial plots covering approximately the outer 270 (of 400 total) radial mesh points in the Vienna pulsation model.
For the phase of maximum radius and the phase of minimum light the gas velocity $u$ (in units of the local sound) speed is plotted.   
At maximum radius (left panel in Fig.\,\ref{comp}) the photosphere starts to move towards the model's center, while 
parts of the envelope still move outwards. All velocities are below sound speed.  
At minimum light (right panel in Fig.\,\ref{comp}) -- which occurs before minimum radius -- we see the transition 
between super- and subsonic inflow, sometimes called a shock, at the photosphere.
This distorts the spectra and makes them less suited for our detailed analysis.

Note that the most quiescent phase, i.e. the phase of maximum radius, is actually very short-lived.  A spectroscopic observation has to be well-timed (within, say, half an hour) in order to catch the spectral lines without distortion. Also, the integration times cannot exceed a few percent of the pulsation period, in order to avoid smearing of the spectra due to pulsation-induced Doppler effect. Integration times not much longer than 15 minutes taken within the appropriate (narrow) phase interval are recommended.  This implies constraints on the obtained signal-to-noise ratio, and therefore, this study could only be done with +2-m telescopes.  At minimum light the star is not at its quietest, and the shock wave associated with the bump phase (Gillet \& Crowe 1988) close to minimum light will also distort the spectral line profiles in RRab stars.

Fig.\,\ref{model-phase} shows clearly that the phase corresponding to minimum light occurs before
a local minimum of the atmospheric kinetic energy, while at the phase of 
maximum radius the stellar atmosphere is very close to the other local minimum of 
kinetic energy.   

\begin{figure}[ht!]
\begin{center}
\includegraphics[width=90mm]{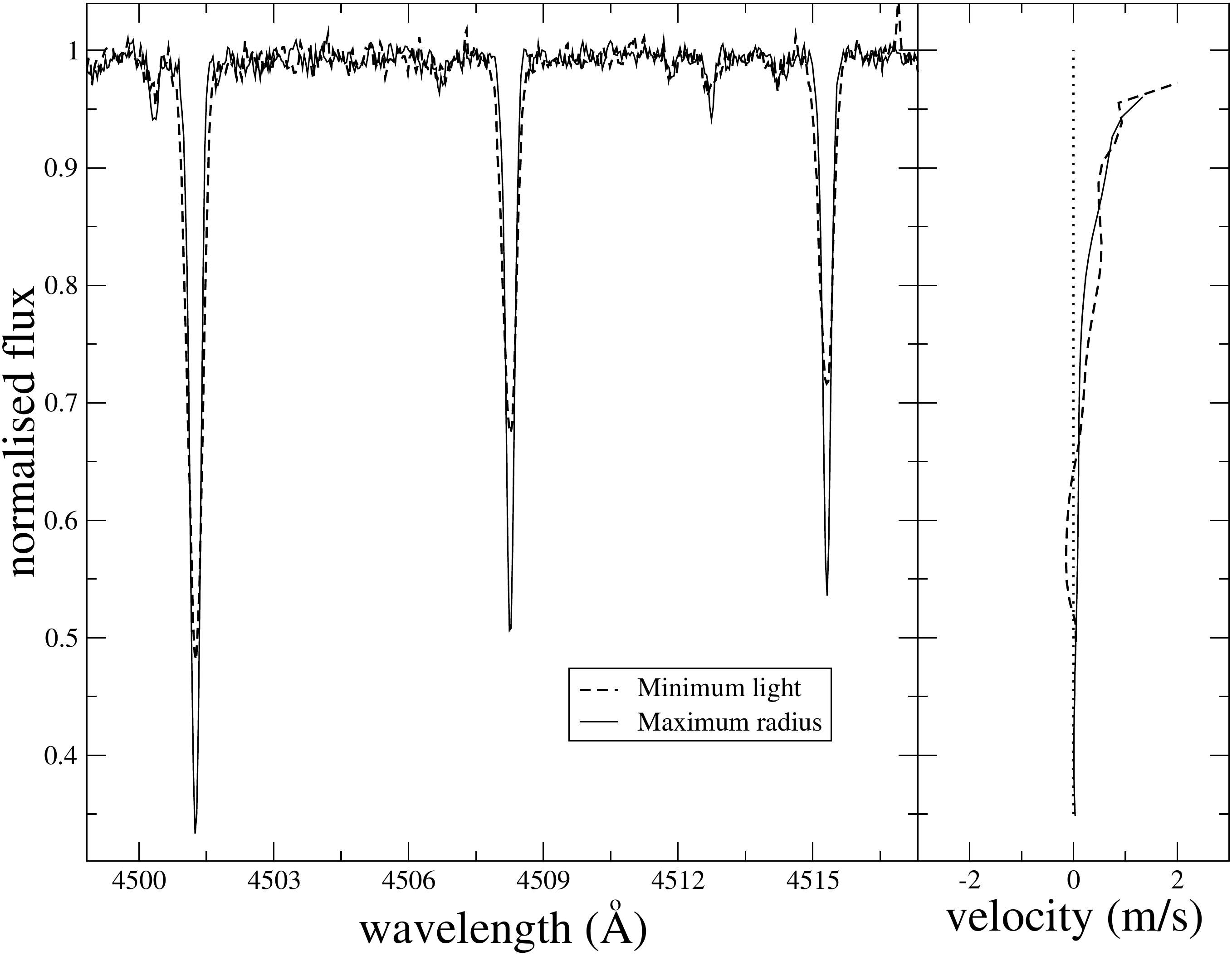}
\caption{Left panel: comparison between a part of the RR~Lyr spectrum as
observed close to the phase of maximum radius (full line) and close to the phase of minimum light
(dashed line). Right panel: comparison between the bisectors of the 
\ion{Ti}{ii} line at $\sim$4501\,\AA\ as observed close to the phase of maximum radius (full line) 
and close to the phase of minimum light (dashed line). The dotted line shows 
the zero velocity, for comparison. The velocity scale is in \ms.}
\label{obs-phase1}
\end{center}
\end{figure}

This picture is also confirmed by the observations. Figure~\ref{obs-phase1} 
shows the comparison of the line profile of RR~Lyr in the region around 
4500\,\AA\ and the bisector of the \ion{Ti}{ii} line at $\sim$4501\,\AA\ as
observed at the phases close to maximum radius and to minimum light. The main difference between the two line profiles is given by the line broadening, which is an indicator of the atmospheric activity: to a quiet phase correspond narrow spectral lines.

\begin{figure}[ht!]
\begin{center}
\includegraphics[width=90mm,clip]{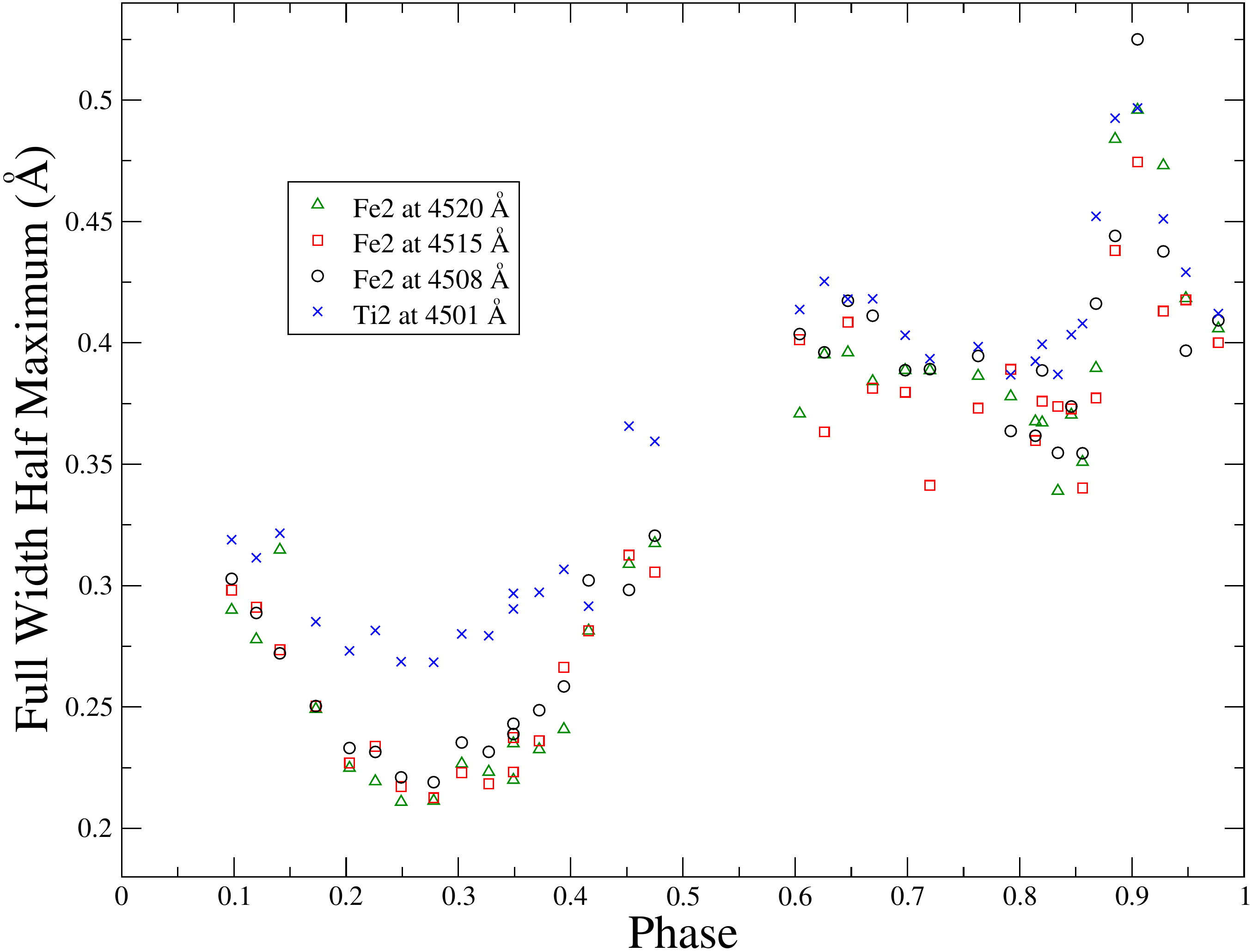}
\caption{Full width at half maximum (FWHM) in \AA\ as a function of pulsational 
phase measured on three \ion{Fe}{ii} lines and one \ion{Ti}{ii} line. The 
minimum is in correspondance with the phase of maximum radius. The typical 
uncertainty on the FWHM is of $\sim$0.01\,\AA, but it depends also on the 
SNR of each spectrum. The two peaks visible at about phase 0.65 and 0.9 are 
due to two propagating shock waves (Fokin et al. 1999).}
\label{obs-phase2}
\end{center}
\end{figure}

The spectral line broadening as a function of phase is shown in 
Fig.~\ref{obs-phase2}. This figure displays the full width at half maximum 
(FWHM) measured for four strong spectral features as a function of the 
pulsational phase. For each of the four lines we obtained a clear
minimum close to the phase of maximum radius (between 0.2 and 0.3). 
This plot also shows  the rapid changes in the FWHM when the shock waves occur. 
The observed peaks in FWHM near pulsation phase $\sim$0.65 and $\sim$0.9 
have been interpreted as arising from two shocks, a weaker and a 
stronger shock respectively, propagating through the star's atmosphere 
and producing compression of the turbulent gas (Fokin et al. 1999). Moreover, 
Fokin \& Gillet (1997) and Fokin et al. (1999) showed that their  RR~Lyr models  exhibit 
very strong shocks\footnote{Fokin \& Gillet (1997)
note that the theoretical velocities and the shock
amplitudes are very sensitive to model parameters. They used parameters different from ours: $T_{\rm eff}$ = 7175 K, $L$ = 62$L_{\odot}$, $M$ = 0.578$M_{\odot}$, $X$ = 0.7 and $Y$ = 0.299. The model generated with the Vienna Nonlinear Pulsation Code shows a maximum, during one pulsation cycle,
of 4.7 Mach outward and  -3.1 Mach inward,
which correspond to gas velocities of  35 and  -23  km/s. 
Due to artificial viscosity we can assume that the star undergoes stronger shocks than our 
models. The artificial tensor viscosity used broadens the shock region and underestimates 
any possible heating phenomena. 
The position of the outer boundary condition is crucial in the maximum velocity, as 
shock waves steepen up when they run outwards towards smaller density. 
Fokin's model runs till $\varrho = 10^{-14}$ while the Vienna models we used stop at about $10^{-10}$.}  up to Mach 25 in the highest part of the star's atmosphere 
(see Fig.~3 in Fokin et al. 1999).

The line bisectors, shown in Figure 3, also display that the
line asymmetry changes with the pulsational phase. In particular,
it has a trend similar to the one shown by the FWHM. This point will be described in 
more detailed in the next paper.

In practice it is not trivial to predict when
the star will be at maximum radius exactly, due to the presence of the Blazhko effect and of 
other (possibly longer) periodicities involved in the pulsation. To find the appropriate 
spectrum for analysis, we investigated all the available spectra obtained close to 
the phase of maximum radius (determined with the simultaneous photometry) 
and picked the ones showing the minimum FWHM and line asymmetry. In the end 
the two spectra with numbers 258 and 260 were obtained very close to the phase of maximum radius and 
showed a comparable line broadening and a very small line asymmetry. 
We decided to perform a detailed 
analysis of spectrum number 260 because of its 
higher SNR. Our simultaneous photometry (Kolenberg et al. 2006) 
confirmed that this spectrum was recorded around pulsation phase $\phi = 0.28$.
\section{Fundamental parameters and abundance analysis}\label{parameters}
In general a fundamental parameter determination begins from a derivation of
\Teff\ and \logg\ from photometric indexes. For RR~Lyr this operation is not trivial. 
In Sect.~\ref{right-phase} we assumed that the atmosphere of RR~Lyr
can be at best simplified as "static" only at the phase of maximum radius.
Therefore it is at this phase that we made use of the photometric indices and the static model atmosphere grids to determine the fundamental parameters.  

As starting point for our analysis we decided to take the parameters derived 
by other authors who analysed spectra of RR~Lyr obtained at a similar phase. 
In particular Takeda et al. (2006) derived the fundamental parameters
spectroscopically from high resolution spectra of RR~Lyr, one of them obtained
not far from the phase of maximum radius. These parameters can be taken only as
starting point because the star was observed at a different Blazhko 
phase\footnote{Fundamental parameters obtained at different Blazhko phases 
are not necessarily equal to one another, although they are obtained at the same pulsation 
phase.}.  We will explore the effect of Blazhko modulation on the spectra of RR~Lyr in our 
forthcoming work.

We used the parameters given by Takeda et al. (2006) for their spectrum taken at pulsation phase 
$\phi=0.36$ (\Teff=6040$\pm$40\,K, \logg=2.09$\pm$0.1\,dex) 
as our starting point.  We performed an iterative process to improve 
and test the parameters as described in the following. In our analysis, every time any of the parameters \Teff, \logg, \vmic, or abundances changed during the iteration process, we 
calculated a new model by implementing the most recently determined quantities.  We did the same with respect to the abundances. While the results of the abundance 
analysis depend upon the assumed model atmosphere, the atmospheric 
temperature-pressure structure itself depends upon the adopted abundances.
We therefore recalculated the model atmosphere every time the abundances were 
changed, even if the other model parameters were kept fixed. This procedure 
ensured that the model structure was consistent with the assumed abundances.

\subsection{The effective temperature}
\begin{figure}[ht!]
\begin{center}
\includegraphics[width=90mm,clip]{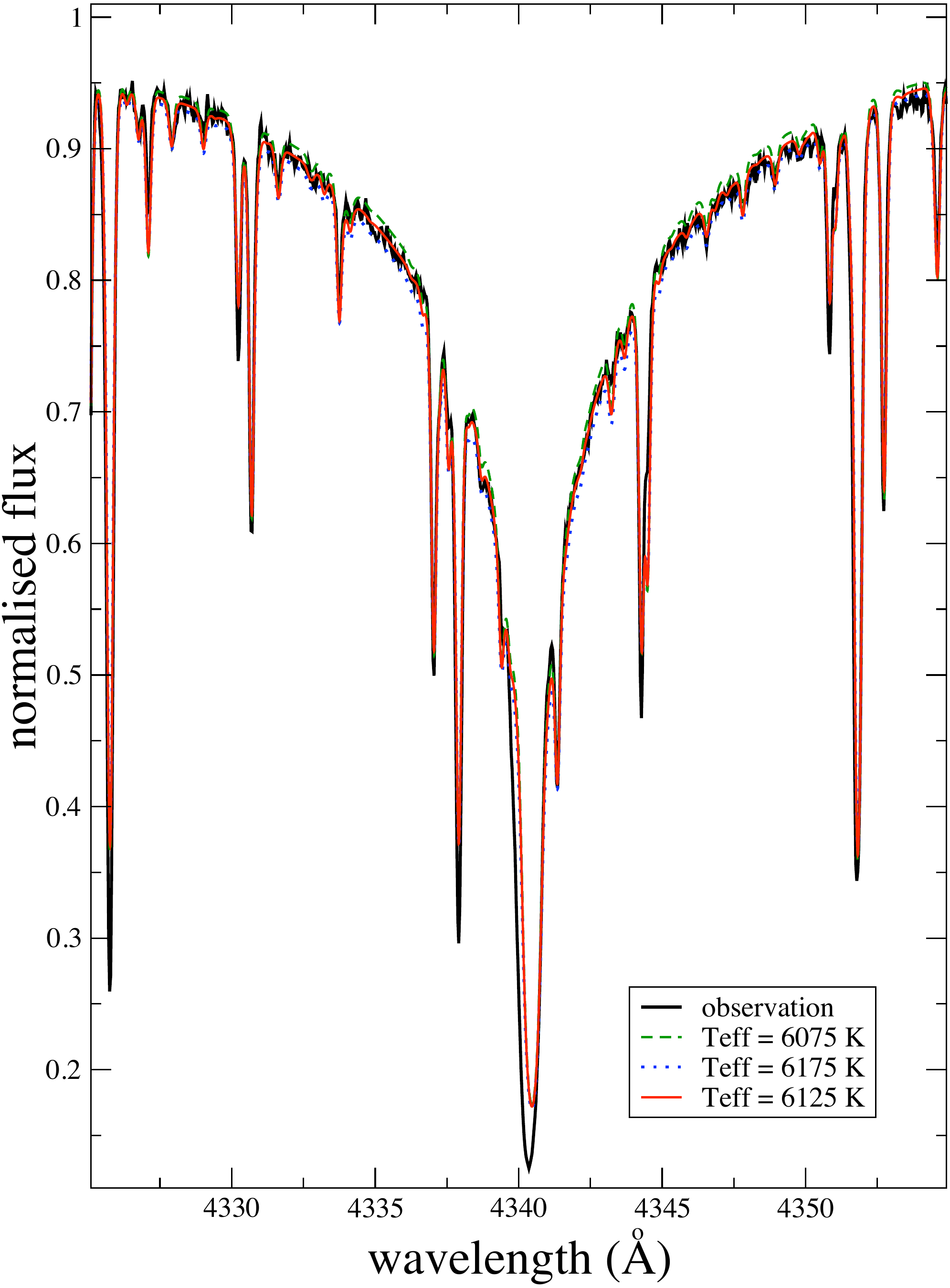}
\caption{Comparison between the observed H$\gamma$ line profile (solid line) 
and synthetic profiles calculated with the final adopted \Teff$=6125$\,K (red
line) and with \Teff=$6175$\,K (green line) and \Teff=$6075$\,K (blue line). 
The red line agrees very well with the observed spectrum.}
\label{hydrogen}
\end{center}
\end{figure}

We performed the \Teff\ determination by fitting synthetic line profiles, 
calculated with \synth\ (Kochukhov 2007), to the observed profile of the 
H$\gamma$ line, the only hydrogen line for which it was possible to make a reliable
normalisation. In the temperature range expected for RR~Lyr, hydrogen lines 
are very sensitive to temperature variations and depend very little on \logg\ 
variations.  In particular, this is expected when the stellar \Teff\ is close 
to its minimum. In the case of RR~Lyr the use of hydrogen lines as \Teff\ 
indicators is very important because these lines describe the stellar structure
more effectively than any other line, being formed in a wide region of the stellar
atmosphere, and the line wings are free from \nlte\ effects. The \Teff\ 
obtained with this procedure is \Teff\ = 6125 $\pm$ 50\,K. 
Note that these error bars are what we obtain from the fitting of the hydrogen line profile.
As there are model uncertainties that are not taken into account in this fitting procedure, these error bars are probably underestimated.
Figure~\ref{hydrogen} shows the comparison between the observed H$\gamma$ 
line profile and the synthetic profiles calculated with the adopted stellar 
parameters, as well as the synthetic profiles obtained increasing and 
decreasing \Teff\ by 50\,K.

Figure~\ref{hydrogen} shows a clear wavelength shift of the observed H$\gamma$ core relative to its wings and to the metal lines formed in deeper layers. This is due to a velocity gradient in the outer layers of the star (Van Hoof \& Struve 1953; Mathias et al. 1995 for RR~Lyr).

Another spectroscopic indicator for \Teff\ is given by the analysis of metallic
lines. In particular, \Teff\ is determined by eliminating the correlation 
between line abundance and line excitation potential (\exc) for a given 
ion/element. This procedure can lead to erroneous parameters, in particular for
stars such as RR~Lyr where \nlte\ effects could be large and the \vmic\ determination is complicated by the pulsation (see Sect.~\ref{discuss-vmic}). For this reason we decided not to take this indicator into account in our analysis, but to use it just as a check of 
the adopted \Teff\ value. This check pointed towards values comparable to the one we found from the H$\gamma$ line profile fitting.

\subsection{The surface gravity}

The surface gravity can be derived using two independent methods:
based on line profile fitting of gravity-sensitive
metal lines with developed wings, and based on
the ionisation balance of several elements.
For RR Lyrae stars the  \ion{Mg}{i} lines are the most suitable lines for the first method. Described in Fuhrmann et al. (1997), the 
first method assumes that the wings of the \ion{Mg}{i} lines at 
$\lambda\lambda$ 5167, 5172, and 5183\,\AA\ are very sensitive to \logg\ variations. 
We decided to use this method only as a check on  \logg\ and not for parameter determination.
The first reason for doing this is the large uncertainty
in the \vsini\ and \vmac\ values that, given the available spectral
resolution and SNR, could not be precisely determined. Another
reason is the slight line asymmetry (clearly visible for the
strong lines, such as \ion{Mg}{i}), which makes precise line profile fitting impossible.
A third reason is the uncertainty in the \vmic\ value, that will be discussed in 
detail in Sect.~\ref{discuss-vmic}.  To make the comparison even more difficult there is the fact that the Mg abundance is about 1\,dex below solar, making the wings of these lines not very pronounced.

The second method for surface gravity determination uses the
assumption of ionisation equilibrium, but this method is extremely sensitive to the \nlte\ 
effects present for each ion/element. Since we are not able to use the 
line profile fitting of the \ion{Mg}{i} lines with developed wings, we rely on the ionisation equilibrium to determine \logg, checking the obtained 
result with the \ion{Mg}{i} lines. In adopting the ionisation equilibrium, for
some elements we took into account also the \nlte\ corrections predicted by
various authors on some specific elements in solar-type metal-poor stars. From the 
ionisation equilibrium we obtained \logg=2.4$\pm$0.2. This value was derived 
using only the lines with an equivalent width of less than 75\,m\AA, in order 
to minimise both the \nlte\ effects and the uncertainty on the \vmic\  (both 
more pronounced for the strong lines), and to assure a large enough number 
of lines, in particular for iron. We checked this value against the observed 
profiles of the \ion{Mg}{i} lines with developed wings and obtained a good 
agreement, in particular when we adopt a depth-dependent \vmic\  
(see Sect.~\ref{discuss-vmic}). For this comparison we used the line 
parameters of the \ion{Mg}{i} lines adopted by Ryabchikova et al. (2009). 

Our value for the surface gravity is supported by the ionisation equilibrium 
for \ion{Fe}{i}/\ion{Fe}{ii} and few other elements, such as 
\ion{Si}{i}/\ion{Si}{ii} and \ion{Ti}{i}/\ion{Ti}{ii}. 
For Ca and V we do not obtain ionisation equilibrium, even within the error
bars, but we have measured just one line for both \ion{Ca}{ii} and \ion{V}{i}.
Taking into account the \nlte\ corrections for \ion{Ca}{i} 
($\sim+$0.1--0.2) and \ion{Ca}{ii} (almost in LTE) given by 
Mashonkina et al. (2006), ionisation equilibrium is achieved for that element.  In the case of chromium, several \ion{Cr}{i} and \ion{Cr}{ii} lines have theoretically calculated oscillator 
strengths, which may influence the final abundance results.  

Since RR~Lyr's effective temperature is too low to show enough
He lines (though they are detected - see Preston 2009), we are unable to measure the atmospheric He abundance. 
Ryabchikova et al. (2009) tested the effect of a strong He depletion in the atmosphere of
the solar type star HD~49933, concluding that a depleted He abundance would
affect only the \logg\ determination and leave  \Teff\  unchanged within the error bars of
0.2~dex.  We also tested  the effect of a He overabundance for RR~Lyr. If we assume $X=0.5$ and $Y=0.49$ we register a general abundance decrease, e.g. Fe decreases 0.3~dex. In addition, we observe a variation of the pressure sensitive features such as the \ion{Mg}{i} lines with extended wings leading to changes in \logg\ that do not exceed our error bars.

\subsection{LTE abundance analysis}
Our main source for the atomic parameters of spectral lines is the \vald\ 
database. LTE abundance analysis was based on equivalent widths, analysed 
with a modified version 
(Tsymbal 1996)
of the WIDTH9 code 
(Kurucz 1993).
We opted for equivalent widths because of the small line asymmetry and of the
uncertainty on the form of the microturbulent velocity, making the synthetic
line profile fitting more uncertain.  We intend to analyse the other collected spectra of RR~Lyr in the 
same consistent way. These spectra
show a much more pronounced line asymmetry, therefore they will be
analysed mostly through equivalent widths. 

In total about 700 lines were measured with equivalent widths, but after a check against both the solar spectrum 
and the spectrum of HD~49933 
(Ryabchikova et al. 2009)
we chose to keep 617 lines of
26 different elements and 32 different ions. We also tried to keep a set of 
lines uniformly distributed over the range of equivalent widths, wavelength, 
and excitation potentials, in particular for important ions such as \ion{Fe}{i} 
for which we kept 284 lines. We used nearly all unblended spectral lines with accurate
atomic parameters, except lines in spectral regions where the
continuum normalisation
was too uncertain.  RR~Lyr shows a strong underabundance for almost every measured 
ion/element.  For this reason, it was possible to give only approximate abundance values or 
upper limits to the abundance of some ions for which it was not possible to 
calculate the equivalent width because of very weak lines. For
these measurements we used synthetic line profile fitting, since these lines
were too shallow to show both any visible line asymmetry and \vmic\ dependence. 

Microturbulence was determined by minimizing the correlation between
equivalent width and abundance for several ions. We used mainly
\ion{Fe}{i} lines since this is the ion that provides the largest number 
of lines within a wide range in equivalent widths, but the
correlations obtained with \ion{Ti}{i}, \ion{Ti}{ii}, \ion{Cr}{i}, 
\ion{Cr}{ii}, \ion{Fe}{ii}, and \ion{Ni}{i} were also taken into 
account. Using all the available lines we could not find a unique 
value for the \vmic\ able to completely remove the correlation between 
equivalent width and abundance.  In particular it was possible to remove 
effectively this correlation using only the lines with an equivalent width 
of less than 75\,m\AA, while the stronger lines exhibited a steep abundance 
increase with increasing equivalent width. Using only the lines with an 
equivalent width of less than 75\,m\AA\ we obtained a \vmic\ of 
2.4$\pm$0.3\,\kms, while using all available lines we obtained a value of 
3.1$\pm$0.5\,\kms.  In this last case the plot shows that
the correlation is only statistically minimised (see lower plot of 
Fig.~\ref{varVmic}). 

In the literature several authors mentioned the 
possibility of a depth-dependent \vmic\ for RR~Lyrae stars (Clementini et al. 1996, Takeda et al. 2006).  Therefore, given the 
impossibility of finding a clear \vmic\ value, we decided to derive the 
profile of a depth-dependent \vmic\ using the available measured lines. 
This part of the work will be discussed in Sect.~\ref{discuss-vmic}. 

The full set of derived abundances, adopting both a constant and a depth
dependent \vmic\, is shown in Table~\ref{abundance}. The last column of 
Table~\ref{abundance} gives the solar abundances by 
Asplund et al. (2005)
for comparison.
\begin{table}[ht!]
\caption[ ]{LTE atmospheric abundances for RR~Lyr.}
\label{abundance}
\begin{center}
\scriptsize{
\begin{tabular}{l|c|c|c|c|c}
\hline
\hline
Ion &\multicolumn{4}{|c|}{RR~Lyr}& Sun \\                    
    &$\log (N/N_{\rm tot})$ & $\log (N/N_{\rm tot})$ & $n$ & Remarks &$\log (N/N_{\rm tot})$  \\
\hline
\ion{C}{i}    & ~~$-$4.84$\pm$0.04 & ~~$-$4.78$\pm$0.07 &  4 &      & ~~$-$3.65~ \\	     
\ion{N}{i}    & ~~$-$4.77$\pm$0.11 & ~~$-$4.77$\pm$0.11 &  2 & S    & ~~$-$4.26~ \\	     
\ion{O}{i}    & ~~$-$3.98$\pm$0.04 & ~~$-$3.97$\pm$0.04 &  6 & S    & ~~$-$3.38~ \\	     
\ion{Na}{i}   & ~~$-$7.38$\pm$0.05 & ~~$-$7.38$\pm$0.05 &  2 &      & ~~$-$5.87~ \\	     
\ion{Mg}{i}   & ~~$-$5.54$\pm$0.10 & ~~$-$5.48$\pm$0.14 &  8 &      & ~~$-$4.51~ \\	     
\ion{Al}{i}   & ~~$-$6.59$\pm$0.26 & ~~$-$6.57$\pm$0.26 &  9 & S    & ~~$-$5.67~ \\	     
\ion{Si}{i}   & ~~$-$5.56$\pm$0.11 & ~~$-$5.55$\pm$0.11 & 10 &      & ~~$-$4.53~ \\	     
\ion{Si}{ii}  & ~~$-$5.61$\pm$0.04 & ~~$-$5.54$\pm$0.08 &  3 &      & ~~$-$4.53~ \\	     
\ion{S}{i}    & ~~$-$5.94$\pm$0.08 & ~~$-$5.93$\pm$0.08 &  7 & S    & ~~$-$4.90~ \\	     
\ion{K}{i}    & ~~$-$7.81          & ~~$-$7.67  	&  1 &      & ~~$-$6.96~ \\	     
\ion{Ca}{i}   & ~~$-$6.98$\pm$0.05 & ~~$-$6.90$\pm$0.09 & 24 &      & ~~$-$5.73~ \\	     
\ion{Ca}{ii}  & ~~$-$6.71    & ~~$-$6.69	&  1 &      & ~~$-$5.73~ \\	     
\ion{Sc}{ii}  & ~$-$10.17$\pm$0.13 & ~$-$10.18$\pm$0.09 & 17 &      & ~~$-$8.99~ \\	     
\ion{Ti}{i}   & ~~$-$8.33$\pm$0.08 & ~~$-$8.32$\pm$0.07 & 22 &      & ~~$-$7.14~ \\	     
\ion{Ti}{ii}  & ~~$-$8.21$\pm$0.19 & ~~$-$8.22$\pm$0.13 & 56 &      & ~~$-$7.14~ \\	     
\ion{V}{i}    & ~~$-$9.64          & ~~$-$9.62  	&  1 &      & ~~$-$8.04~ \\	     
\ion{V}{ii}   & ~~$-$9.22$\pm$0.09 & ~~$-$9.19$\pm$0.09 &  8 &      & ~~$-$8.04~ \\	     
\ion{Cr}{i}   & ~~$-$7.98$\pm$0.09 & ~~$-$7.94$\pm$0.09 & 18 &      & ~~$-$6.40~ \\	     
\ion{Cr}{ii}  & ~~$-$7.70$\pm$0.11 & ~~$-$7.66$\pm$0.10 & 27 &      & ~~$-$6.40~ \\	     
\ion{Mn}{i}   & ~~$-$8.47$\pm$0.15 & ~~$-$8.42$\pm$0.12 &  7 &      & ~~$-$6.65~ \\	     
\ion{Fe}{i}   & ~~$-$6.07$\pm$0.12 & ~~$-$6.03$\pm$0.11 &284 &      & ~~$-$4.59~ \\	     
\ion{Fe}{ii}  & ~~$-$5.93$\pm$0.13 & ~~$-$5.89$\pm$0.10 & 47 &      & ~~$-$4.59~ \\	     
\ion{Co}{i}   & ~~$-$8.22          & ~~$-$8.21  	&  1 &      & ~~$-$7.12~ \\	     
\ion{Ni}{i}   & ~~$-$7.35$\pm$0.08 & ~~$-$7.33$\pm$0.08 & 38 &      & ~~$-$5.81~ \\	     
\ion{Cu}{i}   & ~~$-$9.74          & ~~$-$9.73  	&  1 &      & ~~$-$7.83~ \\	     
\ion{Zn}{i}   & ~~$-$9.01$\pm$0.01 & ~~$-$8.99$\pm$0.01 &  2 &      & ~~$-$7.44~ \\	     
\ion{Ga}{i}   & ~$-$10.50          & ~$-$10.50	 	&  2 & UL/S & ~~$-$9.16~ \\	     
\ion{Rb}{i}   & ~~$-$9.75          & ~~$-$9.75  	&  1 & UL/S & ~~$-$9.44~ \\	     
\ion{Sr}{i}   & ~$-$10.47          & ~$-$10.46  	&  1 &      & ~~$-$9.12~ \\	     
\ion{Sr}{ii}  & ~$-$10.50          & ~$-$10.48  	&  1 &      & ~~$-$9.12~ \\	     
\ion{Y}{ii}   & ~$-$11.31$\pm$0.09 & ~$-$11.30$\pm$0.09 & 10 &      & ~~$-$9.83~ \\	     
\ion{Zr}{ii}  & ~$-$10.56$\pm$0.07 & ~$-$10.53$\pm$0.06 &  4 &      & ~~$-$9.45~ \\	     
\ion{Nb}{ii}  & ~$-$11.50          & ~$-$11.50  	&  1 & UL/S & ~$-$10.62~ \\	     
\ion{Mo}{i}   & ~$-$10.50          & ~$-$10.50  	&  1 & UL/S & ~$-$10.12~ \\
\ion{Pd}{i}   & ~$-$11.00          & ~$-$11.00  	&  1 & UL/S & ~$-$10.35~ \\
\ion{Ba}{ii}  & ~$-$11.23$\pm$0.22 & ~$-$11.35$\pm$0.08 &  5 &      & ~~$-$9.87~ \\	     
\ion{La}{ii}  & ~$-$12.05$\pm$0.02 & ~$-$12.05$\pm$0.02 &  3 &      & ~$-$10.91~ \\	     
\ion{Ce}{ii}  & ~$-$11.74$\pm$0.08 & ~$-$11.73$\pm$0.08 &  8 &      & ~$-$10.46~ \\	     
\ion{Pr}{ii}  & ~$-$12.36$\pm$0.10 & ~$-$12.35$\pm$0.10 & 16 & S    & ~$-$11.33~ \\	     
\ion{Nd}{ii}  & ~$-$11.92          & ~$-$11.91  	&  1 &      & ~$-$10.59~ \\	     
\ion{Sm}{ii}  & ~$-$12.00$\pm$0.07 & ~$-$12.00$\pm$0.07 & 10 & S    & ~$-$11.03~ \\	     
\ion{Eu}{ii}  & ~$-$12.57          & ~$-$12.55  	&  1 &      & ~$-$11.52~ \\	     
\ion{Gd}{ii}  & ~$-$12.00$\pm$0.07 & ~$-$12.00$\pm$0.07 &  5 & S    & ~$-$10.92~ \\	     
\ion{Tb}{ii}  & ~$-$12.25$\pm$0.35 & ~$-$12.25$\pm$0.35 &  2 & S    & ~$-$11.76~ \\	     
\ion{Dy}{ii}  & ~$-$11.82$\pm$0.19 & ~$-$11.80$\pm$0.19 &  6 & S    & ~$-$10.90~ \\	     
\ion{Ho}{ii}  & ~$-$12.75$\pm$0.35 & ~$-$12.75$\pm$0.35 &  2 & S    & ~$-$11.53~ \\	     
\ion{Er}{ii}  & ~$-$11.61          & ~$-$11.60  	&  1 &      & ~$-$11.11~ \\	     
\ion{Tm}{ii}  & ~$-$12.50          & ~$-$12.50  	&  1 & UL/S & ~$-$12.04~ \\	     
\ion{Lu}{ii}  & ~$-$12.83$\pm$0.15 & ~$-$12.83$\pm$0.15 &  3 & S    & ~$-$11.98~ \\	     
\ion{Hf}{ii}  & ~$-$11.50          & ~$-$11.50  	&  1 & UL/S & ~$-$11.16~ \\	     
\ion{Pb}{i}   & ~$-$11.00          & ~$-$11.00  	&  1 & UL/S & ~$-$10.04~ \\	     
\ion{Th}{ii}  & ~$-$12.90          & ~$-$12.90  	&  1 & UL/S & ~$-$11.95~ \\	     
\hline		
\Teff     &\multicolumn{4}{|c|}{6125~K}    & 5777~K  \\ 			    
\logg     &\multicolumn{4}{|c|}{2.40~~~}   & 4.44~~~~\\ 	
\hline		     
\vmic     & 3.1\,\kms & Depth dep. & \multicolumn{2}{c|}{}   & 0.875~~~~\\ 			     
\hline											  
\end{tabular}}
\end{center}
\footnotesize{Error bar are
estimates based on the internal scatter from the number of analysed lines,
$n$. The third column gives the atmospheric abundances in case of a 
polynomial depth dependent microturbulence velocity.
The last column gives the solar abundance values from \citet{met05}. 
The column indicated as "Remarks" shows whether the given abundance value 
is an upper limit (UL) and/or was obtained with synthetic spectra (S).}
\end{table}


The stellar metallicity ($Z$) is defined as follows:
\begin{equation}
\label{Z}
Z_{\rm star}=\frac{\sum_{a \geq 3}m_{a}10^{\log(N_{a}/N_{tot})}}{\sum_{a \geq 1}m_{a}10^{\log(N_{a}/N_{tot})}},
\end{equation}
where $a$ is the atomic number of an element with atomic mass m$_{\rm a}$.
Making use of the abundances obtained from the performed analysis and assuming
the depth-dependent \vmic, we derived a metallicity of 
$Z$ = 0.003 $\pm$ 0.002\,dex. For  elements that were not analysed we adopted solar abundances from Asplund et al. (2005).  However, if we
assume an underabundance of $-1.0$\,dex for all elements that were not analysed,
excluding H and He, the resulting $Z$ value remains practically unchanged.
Note that in the pulsation model we used a metallicity value of $Z$ = 0.001.
This is on the lower border of the value we obtain from our spectrum.  However, we have to keep in mind that we do
not take into account NLTE effects in our analysis, and that these effects may influence our result.
In general, for metal-poor stars such as RR~Lyr, the NLTE correction is negative, meaning that the abundance in NLTE is lower, so the real $Z$ would be lower too (from 0.003 more towards 0.001).  The effect of NLTE on the abundances was illustrated by, e.g., Takeda et al. (2006) for oxygen.
Using the solar Fe abundance value of Asplund et al. (2005), we obtain [Fe/H]=-1.41 $\pm$ 0.11 for RR~Lyr. This is in good agreement with the value [Fe/H]=-1.39 obtained by Beers et al. (2000), who list typical errors of 0.1-0.2 dex.

We used spectral synthesis with \synth\ to check the hyperfine structure (\hfs) effects on the abundance determination of Mn, Cu, Zn, Ba, and Pr. For each measured line of each of these elements \hfs\ effects are less than 0.01\,dex, except for the \ion{Ba}{ii} line at 
$\lambda$~6141\,\AA\ for which the \hfs\ correction is -0.1\,dex, bringing the line abundance closer to the mean Ba abundance. The \hfs\ calculations for barium were taken from McWilliam (1998), who does not list the parameters for the Ba line at $\lambda$~6496\,\AA, but for this specific line we do not expect any significant \hfs\ effect (Mashonkina \& Zhai 2006).

The abundance uncertainties given in Table~\ref{abundance} are the standard
deviation from the mean abundance (hence no uncertainties are given if the abundances were derived from a single line).
More realistic error bars for each
element/ion can be found in Ryabchikova et al. (2009) where a rigorous derivation of the
abundance uncertainties is given on the basis of the adopted uncertainties on
the stellar parameters. This direct comparison is possible because RR~Lyr and
HD~49933 have a similar \Teff\ and in particular similar values of the
uncertainties on both \Teff\ and \logg.

Given the quality of the data and the slight line asymmetry it was not possible
to give definite values for both \vsini\ and \vmac, but just to constrain 
their values. We obtained that \vsini\ lies between 0 and 9\,\kms, and \vmac\ 
between 6 and 11\,\kms. In particular with the minimum given \vsini\ we 
obtained the maximum \vmac\ and vice versa, as both effects contribute to line broadening. Note that the
spectral lines cannot be fit with only the effects of rotational broadening (\vsini).
The constraints on \vsini\ that we obtained are in good agreement with the one obtained by Kolenberg (2002) 
through analysis of the line profile variations of the star.  They also are in accord with the findings by Peterson et al. (1996), 
who measured the line widths for 27 RR Lyrae (of which 8 RRab) variables via cross-correlation analysis. They estimated an upper limit of 
10 km/s for \vsini\  in all cases.

\section{Discussion}\label{discussion}
\subsection{A depth-dependent microturbulent velocity}\label{discuss-vmic}

As previously mentioned, we calculated the profile of a depth-dependent \vmic\ 
on the basis of the available equivalent widths obtained for several \ion{Fe}{i}
lines spanning a large range of values. A depth-dependent \vmic\ was previously 
suggested by several authors, such as
Takeda et al. (2006),
who found evidence that "strongly suggest that the
microturbulence increases with height in the atmosphere of RR~Lyrae stars, and
that a simple application of the \vmic\ value derived from weak/medium-strength
lines to stronger lines may result in an overestimation of the abundances". 

\begin{figure}[ht!]
\begin{center}
\includegraphics[width=90mm,clip]{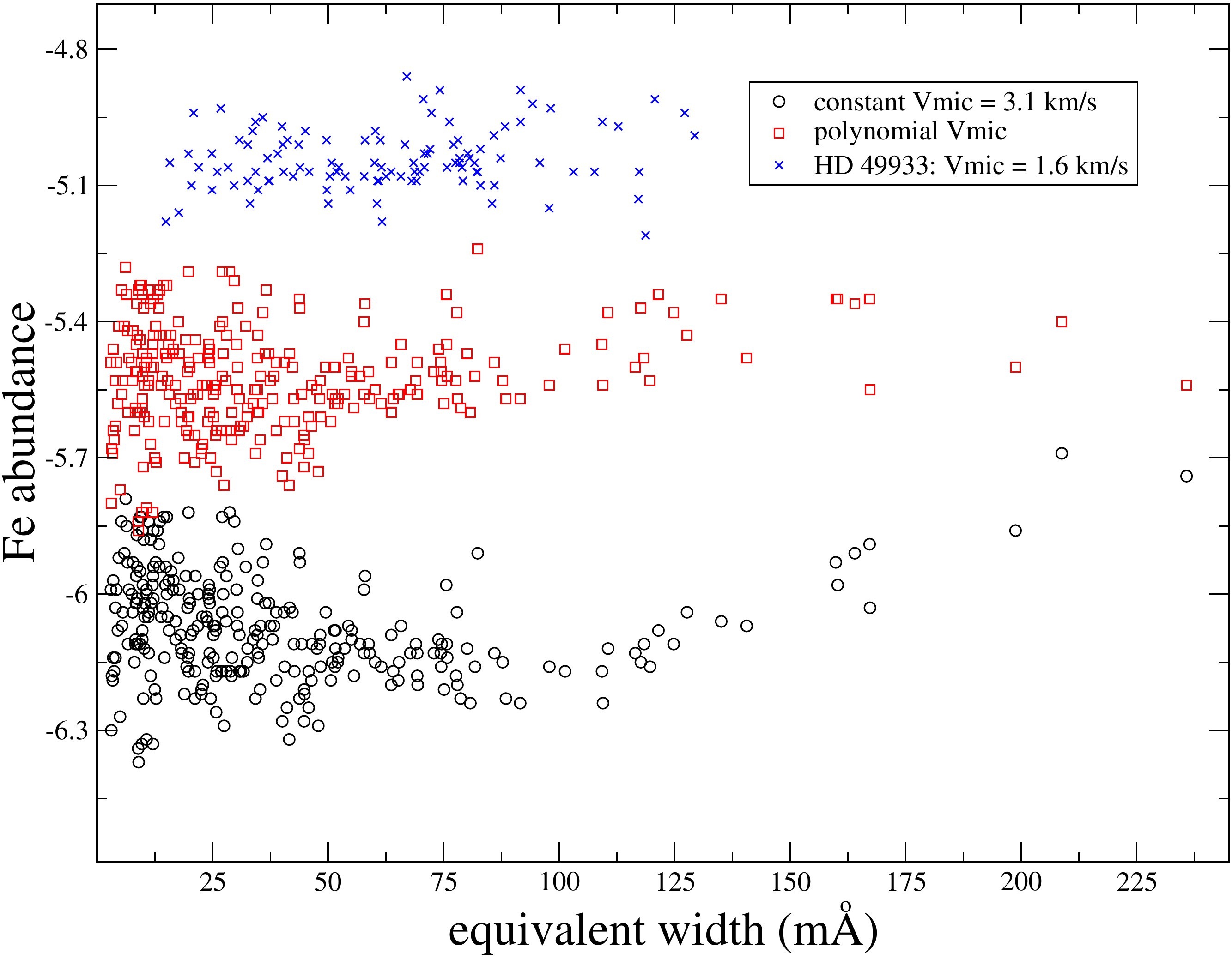}
\caption{Plots of individual abundances for 284 \ion{Fe}{i} lines versus the 
measured equivalent width for RR~Lyr, adopting a constant (open circles) and 
depth-dependent (open squares) \vmic. The same dependence for 108 common 
\ion{Fe}{i} lines in the spectrum of HD~49933 is shown by crosses. We applied an
arbitrary vertical shift for visualisation purposes.}
\label{varVmic}
\end{center}
\end{figure}

Figure~\ref{varVmic} displays the line abundance as a function of the measured
equivalent width for all the measured \ion{Fe}{i} lines in the spectrum of
RR~Lyr, calculated by assuming constant (bottom) and depth-dependent
(middle) \vmic. It is clear that the use of a constant \vmic\ leads to an
underestimation of the abundance of the medium-strength lines and an 
overestimation for the strong lines. In Fig.~\ref{varVmic} we included as
comparison the line abundance as a function of the measured equivalent width  
in HD~49933 for the set of common \ion{Fe}{i} lines. This demonstrates that 
the observed behavior does not depend on the set of adopted lines.

We believe that this phenomenon could be explained both by strong \nlte\ 
effects and by a depth-dependent \vmic\ . Generally speaking, \nlte\ effects 
are stronger for deep lines compared to shallow lines and 
adopting line formation in LTE would lead to a higher abundance, in 
agreement with what we observe here. Gehren et al. (2001) showed that for 
solar-type stars there is a substantial \ion{Fe}{i} underpopulation leading 
to stronger Fe line wings when assumed in LTE, while \ion{Fe}{ii} is in 
LTE, but 
Gehren et al. (2001)
adopted a model atom for Fe that did not include 
high-excitation levels. 
Mashonkina et al. (2009)
analysed the 
\ion{Fe}{i}/\ion{Fe}{ii} ionisation equilibrium in four solar-type stars 
and in the Sun concluding that the inclusion of the high-excitation levels 
in the \ion{Fe}{i} model atom substantially reduced the \nlte\ effects. RR~Lyr 
is a metal poor giant for which \nlte\ effects are expected to be stronger 
than in solar-type stars. If the deviation we register is due to 
\nlte\ effects, it should only produce a deviation of about 0.5\,dex 
for the stronger lines. We believe that the \nlte\ effects are only 
partially responsible for the obtained deviation, because we observe it 
in all measured ions with a similar magnitude and always in the same
direction. These ions include those for which \nlte\ effects are supposed to 
be weak, such as \ion{Fe}{ii}, that shows deviations even stronger than those 
registered for \ion{Fe}{i}. Note that \nlte\ effects work differently for different 
ions, leading to deviations in both directions and with a wide range of 
magnitudes. For this reason we believe that a depth-dependent \vmic\ is mostly
responsible for the observed deviations. A depth-dependent \vmic\ is also 
supported by modelling of RR~Lyr stars as shown by 
Fokin et al. (1999)
who 
observed pulsation-dependent variations in the microturbulent velocity 
(see their Fig.~3).

The depth-dependent \vmic\ profile was obtained fitting\footnote{The code
adopts an LTE plane-parallel model atmosphere calculated using subroutines of
the SynthV code (Tsymbal 1996) and the DUNLSF minimisation procedure of the 
IMSL numerical libraries package.} the line abundance in the plane equivalent
width{ \it versus} line abundance. We performed this procedure for \ion{Fe}{i}, given 
the large number of measured lines and then tested the solution with the other
ions. For each line the code searches the best 
individual line abundance, assuming a certain dependence of the 
microturbulent velocity on the atmospheric depth. This dependence is varied 
to minimise the dispersion between the observed and the theoretical line 
widths over the whole set of measured spectral lines. As the equivalent 
widths come from lines which are formed in a small fraction of the stellar 
atmosphere, it is impossible to obtain a \vmic\ value at each atmospheric 
depth. For this reason it was necessary to speculate about the analytic form 
of the \vmic\ dependence on the atmospheric depth. We tested both a step-like function and a low-degree polynomial function. We chose the latter due to the unrealistically steep and large step needed for the step-like function.

\begin{figure}[ht!]
\begin{center}
\includegraphics[width=90mm,clip]{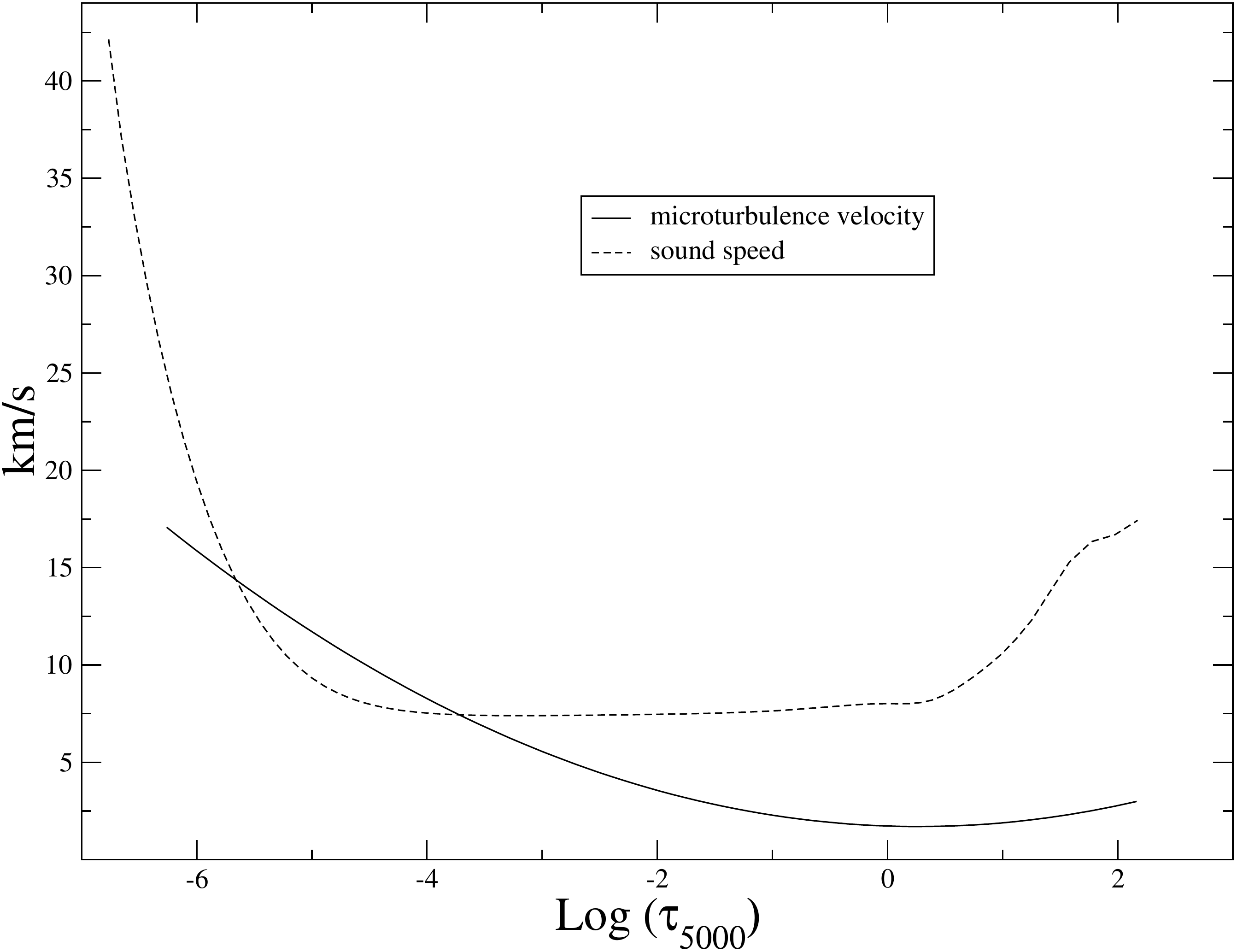}
\caption{Comparison between the profile of the microturbulent velocity (full
line) and of the sound speed (dashed line) as a function of atmospheric 
depth.}
\label{vmic-sspeed}
\end{center}
\end{figure}

Figure~\ref{vmic-sspeed} shows the profile obtained for the depth-dependent
\vmic\ in comparison with the sound speed calculated by \llm. According to the
results of Fokin et al. (1999)
the \vmic\ should always be subsonic, due to the
presence of strong dissipation effects.  In Fig.~\ref{vmic-sspeed} this is the case, except for the region between $\log\tau_{\rm 5000}$ equal to -3.5 and
-5.5 \vmic\ where \vmic\ becomes supersonic. This, however, should be interpreted with caution since
the numeric calculation of the sound speed with the \llm\ code may suffer from accuracy loss in
uppermost layers where the thermodynamic variables (like pressure and density)
are slowly changing functions with atmospheric depth. Also notice that our
empirical estimation of the \vmic\ may contain systematic
uncertainties, and thus the supersonic regime shown in Fig.~\ref{vmic-sspeed}
may have little to do with reality. Nevertheless, the general behavior of 
\vmic\ with depth (i.e., strong increase in superficial layers) 
plauseably reflects a real physical phenomenon. 

Assuming a depth-dependent \vmic\ the abundances of most ions change, in
particular the ones with a large number of strong lines.  As expected, for most of the 
ions the abundance increases instead of
decreasing. This phenomenon is explained by the fact that our
adopted value of constant \vmic\ is overestimated because it was obtained by 
minimising statistically the correlation abundance {\it versus} equivalent width. 
Because we have a large number of middle-strength lines, a small number of
strong lines and an overestimated \vmic\, we obtained an
underestimated abundance (see
the bottom plot of Fig.~\ref{varVmic}). When applying the
depth-dependent \vmic\ , the middle-strength lines (responsible for
the underestimated abundance) yield an abundance similar to the shallow and strong lines, thus
leading to a higher final abundance.
This effect is visible for ions with a large number of
lines, such as \ion{Fe}{i}.
Instead,
for elements with few lines and some of them very strong, such as \ion{Ba}{ii},
the final abundance decreases, with a considerable decrease of the standard
deviation as well.

\subsection{Comparison with previous determinations}\label{comp_param}
Since we propose a different approach to the analysis of RR Lyrae
stars with the aim of determining their physical parameters,
it is important to compare our results with what was previously
obtained by other authors.
The main publications on this topic 
containing an abundance analysis and parameter determination of RR~Lyr itself
are Clementini et al. (1995), Lambert et al. (1996), and Takeda et al. (2006).

Clementini et al. (1995)
analysed several field RR~Lyrae stars to derive broad band
photometric calibrations for both fundamental parameters and metallicity. Their
spectra of RR~Lyr covered a large wavelength range with a moderate resolution
(R$\sim$38\,000) and a high SNR (probably assumed per pixel) of about 460. They
observed RR~Lyr at pulsational phases of 0.70 and 0.73, around minimum
light. Since they did not find any systematic difference between the two spectra
they summed them to get one high-SNR spectrum. They compare the \Teff\ value 
obtained from broad band photometry by several other authors between 1975 
and 1994, and, taking into account these comparisons, they build their own
calibration leading to a \Teff\ value of 6222$\pm$115\,K. They derive the
surface gravity from the stellar mass and radius, obtaining a \logg\ value of 
2.8$\pm$0.2\,dex. The microturbulent velocity was derived in the usual way
(minimisation of the correlation between line abundance and equivalent widths)
obtaining a value of 4.2$\pm$0.2\,\kms. Clementini et al. (1995) also mention the
possiblity of a depth-dependent \vmic, but 
they concluded that, if present, a \vmic\ depth-dependency is only very small 
and affects the abundances by just 0.1\,dex (0.2\,dex for the elements with 
a large number of strong lines). We confirm this estimate with our present work (see Table\,\ref{abundance}).
Most of their abundances are in LTE, exept for
oxygen and sodium that were analysed in \nlte. Table~\ref{comp.clementini} 
shows a comparison between the abundances obtained by Clementini et al. (1995) and 
the ones derived in this work. The comparison shows a rather good general 
agreement between the two sets of abundances, where for only \ion{Al}{i}, \ion{Sc}{ii}, \ion{Mn}{i} and \ion{Zn}{i} we register a small disagreement.
\begin{table}[ht!]
\caption[ ]{Comparison between the atmospheric ion abundances relative to the Sun 
obtained by Clementini et al. (1995) and in this work. }
\label{comp.clementini}
\begin{center}
\scriptsize{
\begin{tabular}{l|c|c|c}
\hline
\hline
Ion &\multicolumn{3}{|c}{[$N_{\rm el}/N_{\rm H}$]$_{\rm Sun}$} \\                    
 & \multicolumn{2}{|c|}{This work} & \citet{clementini} \\
\hline
\ion{O}{i}    & ~$-$0.60$\pm$0.04 & ~$-$0.59$\pm$0.04 & ~$-$0.69          \\	  
\ion{Na}{i}   & ~$-$1.51$\pm$0.05 & ~$-$1.51$\pm$0.05 & ~$-$1.39          \\	  
\ion{Mg}{i}   & ~$-$1.03$\pm$0.10 & ~$-$0.97$\pm$0.14 & ~$-$1.08$\pm$0.07 \\	   
\ion{Al}{i}   & ~$-$0.92$\pm$0.26 & ~$-$0.90$\pm$0.26 & ~$-$1.89          \\	   
\ion{Si}{i}   & ~$-$1.03$\pm$0.11 & ~$-$1.02$\pm$0.11 & ~$-$0.92$\pm$0.02 \\	   
\ion{Si}{ii}  & ~$-$1.08$\pm$0.04 & ~$-$1.01$\pm$0.08 & ~$-$1.14          \\	   
\ion{Ca}{i}   & ~$-$1.25$\pm$0.05 & ~$-$1.17$\pm$0.09 & ~$-$1.07$\pm$0.05 \\	   
\ion{Sc}{ii}  & ~$-$1.18$\pm$0.13 & ~$-$1.19$\pm$0.09 & ~$-$1.35$\pm$0.12 \\	   
\ion{Ti}{i}   & ~$-$1.19$\pm$0.08 & ~$-$1.18$\pm$0.07 & ~$-$1.20$\pm$0.06 \\	   
\ion{Ti}{ii}  & ~$-$1.07$\pm$0.19 & ~$-$1.08$\pm$0.13 & ~$-$1.05$\pm$0.18 \\	   
\ion{Cr}{i}   & ~$-$1.58$\pm$0.09 & ~$-$1.54$\pm$0.09 & ~$-$1.41$\pm$0.18 \\	   
\ion{Cr}{ii}  & ~$-$1.30$\pm$0.11 & ~$-$1.26$\pm$0.10 & ~$-$1.33$\pm$0.14 \\	   
\ion{Mn}{i}   & ~$-$1.82$\pm$0.15 & ~$-$1.77$\pm$0.12 & ~$-$1.99$\pm$0.12 \\	   
\ion{Fe}{i}   & ~$-$1.48$\pm$0.12 & ~$-$1.44$\pm$0.11 & ~$-$1.39$\pm$0.13 \\	   
\ion{Fe}{ii}  & ~$-$1.34$\pm$0.13 & ~$-$1.30$\pm$0.10 & ~$-$1.39$\pm$0.13 \\	   
\ion{Ni}{i}   & ~$-$1.54$\pm$0.08 & ~$-$1.52$\pm$0.08 & ~$-$1.47          \\	   
\ion{Zn}{i}   & ~$-$1.57$\pm$0.01 & ~$-$1.55$\pm$0.01 & ~$-$1.35          \\	   
\ion{Y}{ii}   & ~$-$1.48$\pm$0.09 & ~$-$1.47$\pm$0.09 & ~$-$1.50          \\	   
\ion{Ba}{ii}  & ~$-$1.36$\pm$0.22 & ~$-$1.48$\pm$0.08 & ~$-$1.39$\pm$0.12 \\	     
\hline		
\vmic     & 3.1\,\kms & Depth dep. & ~4.2\,\kms \\ 			     
\hline											  
\end{tabular}}
\end{center}
\footnotesize{The second and third column show the abundances relative to
the Sun obtained in this work, while the fourth column lists the abundances relative to the Sun published by
Clementini et al. (1995).}
\end{table}


Lambert et al. (1996)
observed a set of RR~Lyrae stars with a spectral resolution of
23\,000 to obtain narrow and broad band photometric calibrations for 
fundamental parameters and metallicity. RR~Lyr was observed at eight different 
phases, one of them close to the phase of minimum light and one close to the phase 
of maximum radius. The spectrum obtained close to the phase of maximum radius 
has a SNR (assumed per pixel) of $\sim$100. For each phase they derived the
fundamental parameters both from photometry (adopting previously existing 
calibrations) and from spectroscopy. From the photometric calibration they 
obtained, for the spectrum at phase close to maximum radius, 
\Teff=6350$\pm$200\,K and \logg=2.6$\pm$0.2\,dex, while from spectroscopy 
they derived \Teff=6200$\pm$200\,K and \logg=2.3$\pm$0.2\,dex. Lambert et al. (1996)
found a constant \vmic\ along the stellar atmosphere, but the most surprising 
result is that they obtained also a rather constant \vmic\ along the 
pulsational phase (between 3.6$\pm$0.3\,\kms\ and 4.4$\pm$0.5\,\kms). This 
result is surprising given the fact that several pulsation models of RR~Lyrae 
stars show large variations of the \vmic\ along the pulsational cycle. 
We will study this issue in detail in the forthcoming work. In general the high value obtained 
by Lambert et al. (1996) shows the turbulent motions present in the atmosphere 
and it fits with the large value we also obtained when a constant
\vmic\ is assumed. Lambert et al. (1996) derived the Ca and Fe abundance both in LTE
and \nlte, adopting only the lines that are not deeper than 100\,m\AA. In LTE
they obtained $\log (N_{\rm Fe}/N_{\rm tot})$=-6.06\,dex and 
$\log (N_{\rm Ca}/N_{\rm tot})$=-6.89\,dex. In \nlte\ they obtained a 
correction of about 0.2\,dex for \ion{Fe}{i} and no correction for 
\ion{Fe}{ii}, but, as mentioned in Sect.~\ref{discuss-vmic}, we believe that the
\nlte\ correction for \ion{Fe}{i} is much smaller. The \nlte\ correction they 
obtained for \ion{Ca}{i} was $\sim$0.05\,dex. Both the Fe and Ca
abundance presented by Lambert et al. (1996) are in good agreement with what we
obtained. There is the possibility that the higher \Teff\ and \vmic\ relative 
to what we adopted compensate each other leading to values very close to 
the ones obtained with a lower \Teff\ and \vmic. 

Takeda et al. (2006) analysed five spectra of RR~Lyr with the aim to derive
spectroscopically the fundamental parameters and the abundances of O, Si, and
Fe. The spectra were obtained in high resolution ($R\sim$60\,000) with a rather
high SNR ($\sim$350--400), but in a very limited wavelength range intended to
cover mainly the oxygen triplet at $\lambda\lambda\sim$7770\,\AA. One of these
five spectra was obtained at a pulsational phase close to the phase of maximum
radius. The fundamental parameters were derived minimising the correlation
between line abundance and excitation potential (\Teff) and with the ionisation
equilibrium (\logg).  The hydrogen line wings were used as check of
the parameters determined, the inverse of the strategy we applied.  For this phase
they obtained \Teff=6040$\pm$40\,K, \logg=2.09$\pm$0.1\,dex, and
\vmic=3.0$\pm$0.1\kms. The \vmic\ value obtained by Takeda et al. (2006) is in 
very good agreement with what we have found using a constant \vmic. They 
also strongly suggest the presence of a depth-dependent \vmic\ for RR~Lyrae 
stars. Takeda et al. (2006) analysed two sets of oxygen lines in \nlte\ obtaining two values 
for the O abundance of -4.11\,dex and -4.00\,dex. 
We believe that the difference between their two obtained values is due to the fact that they adopted a constant \vmic.
The strong infrared oxygen triplet is very sensitive to the adopted \vmic.
The O abundance we obtained 
is very close to -4.00\,dex assuming both a constant and a depth-dependent 
\vmic.  
Takeda et al. (2006) also derived the Si and Fe abundance in LTE, obtaining
respectively -5.93\,dex and -5.82\,dex. These two values do not match very well
our results (we obtain a higher Si abundance and a lower Fe abundance).  We
are not able to explain this difference since it cannot be due to the
small differences in the adopted stellar parameters. 

The effective temperatures determined in these studies may not be directly comparable because the analysed spectra were obtained at different Blazhko phases
A more reasonable comparison can be done for the surface gravity because it is supposed to
change less during the pulsation cycle. 

What clearly emerges is a systematic
difference between the \logg\ obtained from stellar mass and radius and from the
ionisation equilibrium. The ionisation equilibrium leads to a lower \logg\ 
compared to what was derived by the assumed stellar mass and radius. Our \logg\ 
value lies in between and it is well validated from the fit of the magnesium 
lines with developed wings. This method is independent of both the ionisation 
equilibrium and the assumed stellar mass and radius. 

The other parameters that
can be directly compared are the elemental abundances.  We do not expect those to 
vary within the pulsation cycle, unless the pulsation is able to bring
material from the inner core up to the stellar atmosphere, a possibility that we find
unlikely. Comparing the abundances obtained in the three previous studies and in our work, we note good agreement except for a few exceptions. This
means that there is the actual possibility that if static models can be applied
to RR~Lyrae stars, they can be applied to many phases (not all!), but we will tackle this
issue in our next work on the analysis of the other acquired spectra of RR~Lyr.

\begin{figure}[ht!]
\begin{center}
\includegraphics[width=\hsize,clip]{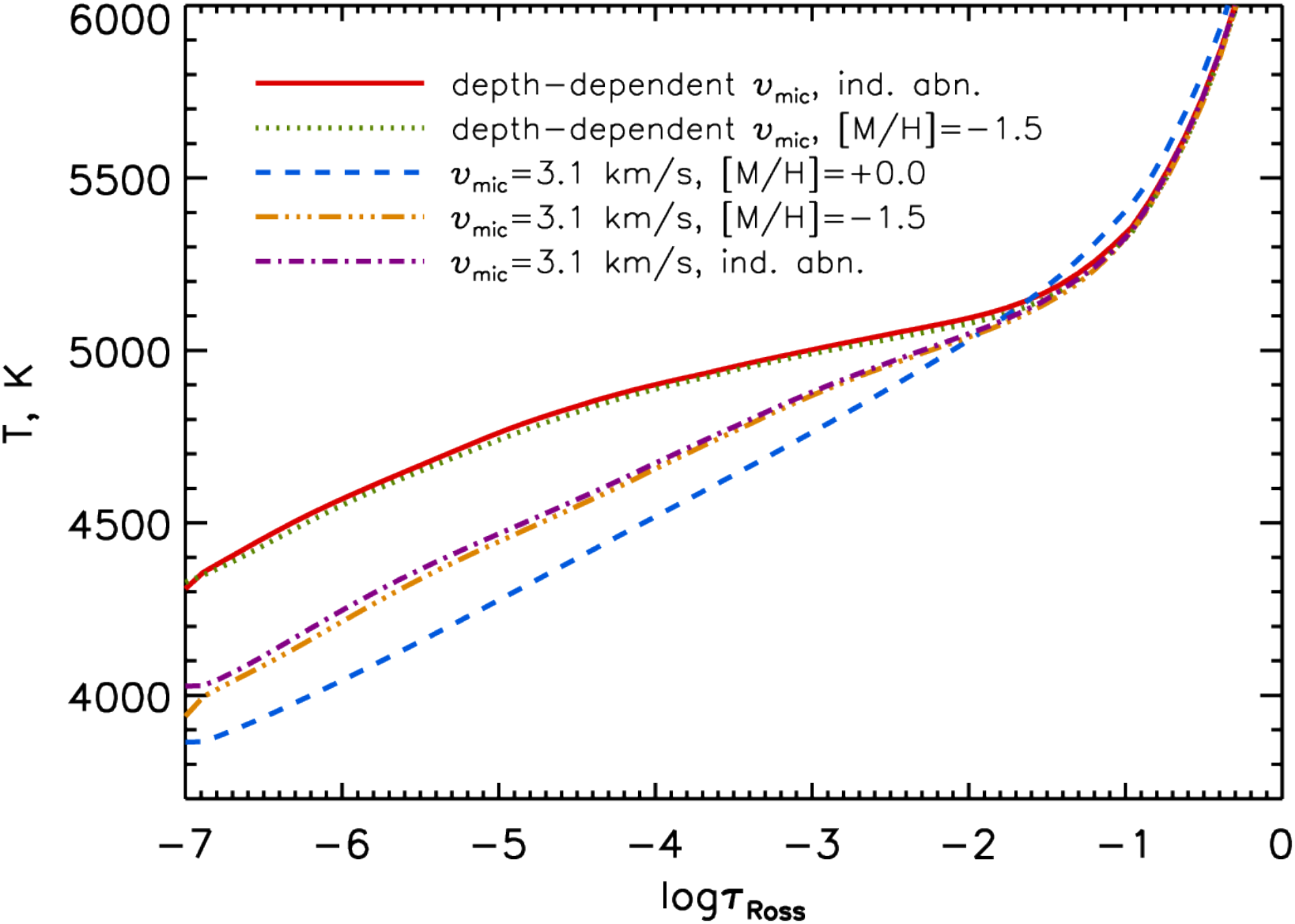}
\caption{Comparison of temperature structures of models computed under 
different assumptions about microturbulent velocity \vmic\
and abundances (see the legend on the plot). Depth-dependent \vmic\ was
taken from Fig.~\ref{vmic-sspeed}. For all the models we took \Teff$=6125$, \logg$=2.4$.}
\label{fig:t}
\end{center}
\end{figure}

From the point of view of the atmospheric modelling, the main difference between earlier work and ours is the adoption of an abundance-dependent model atmosphere for an RR Lyrae type star. The impact of depth-dependent microturbulent velocity
and individual chemistry is shown in Fig.~\ref{fig:t}. As expected, strong
underabundance makes the temperature in the surface layers higher
compared to solar or scaled solar models. Introduction of the depth-dependent
\vmic\ affects the strongest spectral features, leading to weaker absorption at their line center. This, in turn, decreases the line absorption coefficient and thus leads to a further increase
of the temperature. For instance, the temperature difference between models computed with constant and depth-dependent
microturbulent velocities (see Fig. 7) can amount to as much as 300 K.
On the other hand, only the strongest lines feel this modification in model temperature structure
and thus the statistical results of abundance analysis are not affected that much, as shown in Table\,2. 

In contrast, line profile analysis of strong lines must be performed with appropriate models
that account for the effect of depth-dependent microturbulence. Finally, we do not find any
noticeable difference in temperature structure between the scaled-solar model ([M/H]=-1.5), which
corresponds to the mean underabundance we found for Fe, 
and individual abundance models. This allows us to use scaled abundance
models to mimic the temperature structure of the star
as a first guess. Note however that metallicity is the result of
abundance analysis, which is not known in the beginning. This
methodological difficulty is automatically removed using an iterative procedure
of abundance analysis as applied in this study.

\section{Conclusions}

Fundamental-mode RR Lyrae (RRab) stars pulsate with high velocities which, in certain phases, leads to a distortion of their spectral features.
The aim of the study presented in this paper was to determine a set of self-consistent and accurate parameters and abundances for RR~Lyr, the prototype and brightest member of its class.
For this, we had a set of high-resolution spectroscopic data at our disposal, obtained with the Robert G. Tull Coud\'e
Spectrograph on the 2.7-m telescope of McDonald Observatory.
In order to derive the most reliable abundances for the star, we determined the phase in RR~Lyr's pulsation cycle at which the atmosphere is ``at its most quiescent", the phase of maximum radius. Our assumptions were strengthened by the Vienna Nonlinear Pulsation Code.
With observations taken at the phase of maximum radius, the fundamental parameters and the element abundances of RR~Lyr were determined through an iterative process. 
For the determination of the effective temperature, we used synthetic line profile fitting to the H$\gamma$ line and obtained \Teff\ = 6125 $\pm$ 50 K (error bars from the fitting procedure only).  For the \logg\ determination we used the condition of ionisation equilibrium and obtained a \logg\ value of 2.4 $\pm$ 0.2 at the phase of maximum radius.  
The LTE abundance analysis based on element abundances {\it versus} equivalent widths could not fit the element abundances with a single microturbulent velocity (\vmic\ ) value. A depth-dependent \vmic\ is physically plausible for RR Lyrae atmospheres and was previously suggested (but never derived or quantified) by several authors.
In this work, we derived, for the first time, the depth-dependent \vmic\ profile and 
quantified the expected abundance variation.  With the depth-dependent \vmic\ we obtained a better agreement between the element abundances and equivalent widths.  In general, the adoption of a fixed \vmic\ value (which is too high) leads to an underestimation of the element abundances by 0.06 dex (maximum value).  Nevertheless, RR~Lyr is shown to be underabundant in all heavy elements, in agreement with previous studies.

RR~Lyr, the eponym of its class, is one of the best studied RR~Lyrae stars.  However, many intricacies of its pulsation remain poorly understood, and in order to accurately model the star, we have to take into account complex physics that we are only beginning to uncover. 
The star is an asteroseismic target of the {\it Kepler} Mission (Borucki et al. 2009) through the Kepler Asteroseismic Science Consortium (KASC, see also Gilliland et al. 2010). Besides the fact that the star displays amplitude and phase modulation (the so-called Blazhko effect),
Kolenberg et al. (2010) recently detected, for the first time, the occurrence of half-integer frequencies in the star, i.e., peaks at $\frac{1}{2} f_0, \frac{3}{2} f_0, \frac{5}{2} f_0$ etc., with $f_0$ the main pulsation frequency . This phenomenon (``period doubling") may be caused by instabilities in the star and connected to the mysterious Blazhko effect (see Szab\'o et al. 2010, in preparation).
In order to further explore theoretical models of RR~Lyr, it is necessary to know the physical parameters of the star with the highest accuracy possible. 

For this reason, an in-depth analysis of the star's atmospheric motions is very timely.
We intend to expand this analysis in our forthcoming publications of the star.
Our study clearly illustrates that it is crucial to use the appropriate models to correctly interpret the spectral data of RR~Lyrae stars, and that high-quality observations can contribute to improving those models.

\begin{acknowledgements}
We kindly thank the referee of this paper, Dr. George Preston, for constructive comments.
KK is a Hertha Firnberg Fellow, supported by the Austrian Science Foundation (FWF project T359-N2 and FWF stand-alone project P19962). LF has received support from the Austrian Science Foundation (FWF project P19962). His research at the Open University (UK) is funded by an STFC Rolling Grant. DS is supported by  Deutsche Forschungsgemeinschaft (DFG) Research Grant RE1664/7-1. 
OK is a Royal Swedish Academy of Sciences Research Fellow supported by grants from the Knut and Alice Wallenberg Foundation and the Swedish Research Council. 
We kindly thank Stefano Bagnulo and John Landstreet for the useful discussions and comments during the preparation of the draft.
\end{acknowledgements}

\end{document}